\documentclass[%
 reprint, prd,
superscriptaddress,
nofootinbib,
 amsmath,amssymb,
 aps,
]{revtex4-1}
\usepackage{xcolor}
\usepackage{graphicx}
\usepackage{dcolumn}
\usepackage{booktabs}
\usepackage{bm}
\usepackage[dvipsnames]{xcolor}

\usepackage[normalem]{ulem}

\usepackage{rotating} 

\usepackage{url}
\usepackage[linktocpage]{hyperref}
\usepackage{booktabs} 

\hypersetup{
	colorlinks,
	linkcolor={red!60!black},
	citecolor={green!50!black},
	urlcolor={blue!80!black}
}

\usepackage{array} 

\usepackage{orcidlink}
\usepackage{fontawesome5} 
\definecolor{colorgit}{rgb}
{0.30,0.52,0.62}

\newcommand{\gitlinkcato}{%
  \href{https://github.com/davidcato/class-interacting-neutrinos-PT}{%
    \nolinkurl{https://github.com/davidcato/class-interacting-nu}} %
  {\large\color{colorgit}\faGithub}%
}
\newcommand{\gitlinkfolps}{\url{https://github.com/cosmodesi/FolpsD} {\large\color{colorgit}\faGithub}}
\newcommand{\gitlinkharmonic}{\url{https://github.com/astro-informatics/harmonic} {\large\color{colorgit}\faGithub}}

\newcommand{\gitlinkpocomc}{\url{https://github.com/minaskar/pocomc} {\large\color{colorgit}\faGithub}}

\newcommand{\gitlinklyalpha}{%
  \url{https://github.com/igmhub/cobaya_lya_p1d} {\large\color{colorgit}\faGithub}%
  }%

\definecolor{amethyst}{rgb}{0.6, 0.4, 0.8}

\usepackage{soul}
\usepackage[normalem]{ulem}

\usepackage{bm}

\newcommand{\kmsMpc}{\,{\rm km\,s^{-1}\,Mpc^{-1}}}

\newcommand{\ob}{\omega_\mathrm{b}}
\newcommand{\ocdm}{\omega_\mathrm{cdm}}

\newcommand{\sumnu}{\sum m_{\nu}}

\begin{document}

\preprint{APS/123-QED}

\title[Neutrino self-interactions in light of DESI and CMB-SPA]{Neutrino self-interactions in light of DESI and CMB-SPA}

    \author{{Hernán E. Noriega}\orcidlink{0000-0002-3397-3998}}\email{hnoriega@umich.edu}
    \affiliation{Leinweber Institute for Theoretical Physics, University of Michigan, Ann Arbor, MI 48109, USA}
    \affiliation{Department of Physics, University of Michigan, Ann Arbor, MI 48109, USA}

    \author{{Josue De-Santiago}\orcidlink{0000-0002-1163-3730}}\email{Josue.desantiago@cinvestav.mx}
    \affiliation{ Departamento de F\'isica, Centro de Investigaci\'on y de Estudios Avanzados del Instituto Politécnico Nacional.  
                    Apartado Postal 14-740, 07000, Ciudad de M\'exico, M\'exico}
    \affiliation{Secretar\'ia de Ciencia, Humanidades, Tecnolog\'ia e Innovaci\'on,
    			Av. Insurgentes Sur 1582, 03940, Ciudad de M\'exico, M\'exico}

    \author{\mbox{Gabriela~Garcia-Arroyo}\orcidlink{0000-0002-0599-7036}}\email{arroyo@icf.unam.mx}
    \affiliation{Instituto de Ciencias F\'isicas, Universidad Nacional Aut\'onoma de M\'exico, Apartado Postal 48-3, 62251 Cuernavaca, Morelos, M\'exico}

    \author{{Ivan P\'erez-Castro}\orcidlink{0009-0003-1060-6858}}\email{ivan.perez.c@.cinvestav.mx}
    \affiliation{ Departamento de F\'isica, Centro de Investigaci\'on y de Estudios Avanzados del Instituto Politécnico Nacional. 
                    Apartado Postal 14-740, 07000, Ciudad de M\'exico, M\'exico}

   \author{{Jorge Venzor}\orcidlink{0000-0003-4510-2089}}\email{jorge.venzor@cinvestav.mx}
    \affiliation{ Departamento de F\'isica, Centro de Investigaci\'on y de Estudios Avanzados del Instituto Politécnico Nacional. 
                    Apartado Postal 14-740, 07000, Ciudad de M\'exico, M\'exico}
    		   



\begin{abstract}
We present an updated analysis of non-standard neutrino self-interactions, where the interaction is effectively mediated by a \emph{heavy particle}, using the latest available data from DESI and the CMB.
The considered data include combinations of DESI DR1 full-shape galaxy clustering (FS), BAO distance measurements, Lyman-$\alpha$ 1D power spectra (Ly-$\alpha$), and a combined CMB likelihood (CMB-SPA) using data from Planck, ACT, and SPT.
First, we probe two DESI DR1-only datasets, where the effective interaction parameter $G_{\rm eff}$ is multimodal in the profile likelihood, leading us to analyze two different regimes: moderately interacting (MI$_\nu$) and strongly interacting (SI$_\nu$) neutrinos.
With FS+BAO, the evidence disfavors the SI$_\nu$ mode, a trend that becomes slightly stronger with the inclusion of Ly-$\alpha$.
However, in the latter dataset, the MI$_\nu$ region shows a bimodality in which the intermediate mode is very sharp and peaks around $G_{\rm eff\,} \simeq 10^{-3} \mathrm{MeV}^{-2}$; interestingly, this mode is not penalized versus the weaker standard-like mode.
Furthermore, we test CMB and DESI joint data and show that the SI$_\nu$ mode is more disfavored by CMB-SPA than by Planck-only data.
In particular, the evidence against this mode becomes the strongest thus far (with odds of 5:100000) in the FS+BAO+CMB-SPA combination.
\end{abstract}

\maketitle


\section{\label{sec:introduction} Introduction}
Neutrinos are among the least understood particles in the Standard Model. The observation of neutrino flavor oscillations establishes that at least two neutrino mass eigenstates have nonzero masses~\cite{Fukuda1998evidence, Ahmad2002direct}. The measured mass-squared splittings imply lower bounds on the total neutrino mass, $\sumnu\gtrsim0.06\,\mathrm{eV}$ for normal ordering (NO) and $\sumnu\gtrsim0.10\,\mathrm{eV}$ for inverted ordering (IO)~\cite{Esteban2024nufit, Juno2026measurement}. The absolute neutrino mass scale remains undetermined, with direct laboratory constraints provided by beta-decay experiments~\cite{Katrin2025direct}.

Although neutrinos have tiny masses, they are the second most abundant Standard Model particles in the Universe leaving observable cosmological signatures that depend on their masses.
In particular, neutrinos influence cosmology in two main ways. First, they affect the background evolution of the Universe, contributing to the radiation energy density while relativistic and subsequently behaving as matter after becoming nonrelativistic, thereby modifying the overall expansion history. Second, even after becoming nonrelativistic, they retain sufficiently large thermal velocities to inhibit clustering on scales smaller than their characteristic free-streaming scales, suppressing the growth of matter perturbations~\cite{Lesgourgues:2006nd, Wong:2011ip}.

Beyond these mass-dependent effects, cosmological observables have also been shown to probe the free-streaming nature of neutrinos~\cite{Bashinsky2004neutrino, Follin:2015hya, Baumann2019first, Montefalcone2025free}. In the standard thermal history, Standard Model weak interactions become inefficient at temperatures of order ${\rm MeV}$, 
allowing neutrinos to decouple from the primordial plasma and subsequently free-stream~\cite{Dolgov2002neutrinos, Akita2020precision, Froustey2020neutrino, Bennett_2021}. While relativistic, their free-streaming perturbations develop anisotropic stress and modify the gravitational potentials that drive acoustic oscillations, leaving characteristic signatures in the amplitudes and phases of the cosmic microwave background (CMB) acoustic peaks~\cite{Bashinsky2004neutrino}. Sufficiently strong nonstandard neutrino self-interactions (NSI) can suppress neutrino anisotropic stress and delay the onset of free-streaming; see Ref.~\cite{Berryman2023neutrino} for a review.
The resulting changes in the evolution of cosmological perturbations affect both CMB anisotropies and the matter power spectrum, making CMB and large-scale structure (LSS) observations complementary probes of new interactions in the neutrino sector~\cite{Kreisch2020, Camarena2023confronting, He2024self, Camarena2025strong, Poudou:2025qcx, Whitford:2025dmq}.

With the increasing precision of cosmological observations and the growing amount of available data, this scenario has attracted renewed attention because several cosmological analyses exhibit a bimodal posterior distribution, which has motivated further investigation of possible NSI~\cite{Lancaster2017, Oldengott2017, Kreisch2020, Park2019, Barenboim2019inflation, Das2021, Choudhury2021, Brinckmann2020self, Mazumdar2022flavour, Kumar_2022, Kreisch:2022zxp, Das:2023npl, Camarena2023confronting, He2024self, Camarena2025strong, Poudou:2025qcx, He:2025jwp, Calabrese2025atacama}.
The NSI bimodality is commonly associated with two distinct interaction regimes: a moderately interacting $({\rm MI}_\nu)$ mode, whose cosmological predictions remain close to those of standard free-streaming neutrinos, and a strongly interacting $({\rm SI}_\nu$) mode, in which the onset of free streaming can be delayed until near matter--radiation equality~\cite{CyrRacine2014, Archidiacono2014, Lancaster2017, Kreisch2020, Camarena2023confronting}. The relative statistical support for these modes depends on the observations included in the analysis. 
Earlier studies have found a preference for the ${\rm SI}_\nu$ mode~\cite{Park2019, Kreisch2020, Kreisch:2022zxp},
which is weakened by the inclusion of polarization data~\cite{Brinckmann2020self, Choudhury2021, Kreisch:2022zxp, Das:2023npl}. More recent analyses using data from the Atacama Cosmology Telescope (ACT) DR6 find no significant evidence for NSI~\cite{Calabrese2025atacama}. Constraints from galaxy clustering also depend on the combination with CMB data and the treatment of nonlinear clustering~\cite{Camarena2023confronting, Camarena2025strong, Poudou:2025qcx}. Earlier combined analyses including Ly-$\alpha$ forest measurements reported strong support for the ${\rm SI}_\nu$ mode~\cite{He2024self}, but updated analyses of the eBOSS forest using effective field theory (EFT)-based and PRIYA likelihoods favor negligible self-interactions~\cite{He:2025jwp}. 
These results motivate reassessing the viability of strong NSI and the robustness of the inferred preference between the two modes.

The availability of new CMB and LSS measurements provides an opportunity to address these questions. The Dark Energy Spectroscopic Instrument (DESI) supplies galaxy full-shape measurements that probe scale-dependent features induced by nonstandard neutrino self-interactions, together with baryon acoustic oscillation (BAO) measurements that constrain the expansion history~\cite{DESI:2024jis, adame2024desiBAO, Abdul2025desi}. Measurements of the one-dimensional Ly-$\alpha$ forest flux power spectrum extend this sensitivity to smaller scales and higher redshifts~\cite{Chaves-Montero_2026_DESIlya}. On the CMB side, improved temperature and polarization measurements from ACT DR6~\cite{Louis_2025_ACT_DR6} and the SPT-3 D1 dataset from the South Pole Telescope (SPT)~\cite{SPT_2026_CMBSPA}, combined with Planck observations~\cite{Planck:2019nip}, supply information across complementary angular scales. By probing the effects of delayed neutrino free streaming on both CMB anisotropies and matter clustering, the new observations allow a more comprehensive assessment of the two interaction regimes. Recent combinations of DESI and CMB data also yield tight constraints on $\sumnu$ within the flat $\Lambda$CDM model, in conflict with the minimum bounds allowed by neutrino oscillations~\cite{Naredo2024critical, Elbers2025constraints, Calabrese2025atacama, DESI:2025gwf, SPT_2026_CMBSPA}. This further motivates examining how cosmological neutrino constraints depend on assumptions about the neutrino sector.

In this work, we present an updated cosmological analysis of NSI mediated by a heavy particle, using the latest publicly available LSS and CMB observations. We first analyze DESI DR1 galaxy full-shape, BAO and Ly-$\alpha$ measurements with minimal external priors, keeping the analysis CMB-agnostic. We then combine the joint Planck, ACT DR6, and SPT-3 D1 dataset (CMB-SPA) with either DESI DR2 BAO or DESI DR1 FS+BAO measurements. We characterize the multimodal structure of $G_{\rm eff}$, isolate the individual interaction modes, and compare their statistical support using Bayesian evidence, maximum likelihood ratio, and profile likelihood analyses. We also examine the dependence of these results on prior choices and the neutrino parameters $\sumnu$ and $N_{\rm eff}$. By comparing LSS constraints with joint LSS and CMB results, this work takes a step toward clarifying the cosmological status of NSI.

The rest of the paper is organized as follows. In Sec.~\ref{sec:lperturbations}, we review the adopted NSI model. Sec.~\ref{sec:data} describes the datasets and analysis setup. We present the main results in Secs.~\ref{sec:results_lss} and \ref{sec:results_cmb}, and compare the interaction modes using Bayesian evidence and maximum-likelihood ratios in Sec.~\ref{sec:bayesian}. Finally, Sec.~\ref{sec:conclusions} presents our conclusions and outlook.

\section{\label{sec:lperturbations}
Neutrino Self-Interactions in Cosmology}
To explore the effects of NSI on the gravitational potentials and anisotropic stress, we choose a reference neutrino species. We then construct the Boltzmann hierarchy equations (BHEs) from perturbations around the background neutrino distribution function, $f_\nu^{(0)}$.
Working in Fourier space, we can then derive the BHEs in terms of Legendre multipoles of the perturbation~\cite{Ma1995}, with an additional nonvanishing collision term at first order (see, for instance, the implementation using the Maxwell--Boltzmann approximation for the distribution functions in~\cite{Oldengott2015, Oldengott2017, Mogensen2026self}).

Since neutrinos remain relativistic prior to the matter-radiation equality epoch, one can assume that the Legendre multipoles of the perturbation are related to a temperature fluctuation. This allows the momentum
dependence to be integrated out, reducing the hierarchy to a set of momentum-independent neutrino multipoles, $\nu_\ell$, and thereby considerably simplifying the BHEs. This procedure is commonly referred to as the \emph{thermal approximation} and provides a convenient framework for describing massive neutrinos whose free streaming is delayed by NSI.

In the following, we adopt the implementation of~\cite{Kreisch2020} that describes NSI between massive neutrinos mediated by a scalar particle, for which the BHEs in the conformal Newtonian gauge take the expression,
    \begin{equation}
    \begin{split}
        \frac{\partial \nu_{\ell}}{\partial \tau} &+ \frac{q}{\epsilon} k \left[ \frac{\ell +1}{2 \ell +1} \nu_{\ell +1} - \frac{\ell}{2\ell +1} \nu_{\ell -1} \right] \\
        &- 4 f_{g,\ell} = - a G_{{\rm eff}}^2 T_\nu^5 H_{\nu,\ell}(\epsilon) \nu_\ell \, , 
    \end{split} \label{Eq: BHE_CN}
    \end{equation}
where $\tau$ denotes conformal time, $k$ is the comoving wavenumber, $q=ap_1$ is the magnitude of the comoving momentum of the reference neutrino, $\epsilon=\sqrt{q^2+a^2m_\nu^2}$ is its comoving energy, and $m_\nu$ stands for the mass of the neutrino species. The term $f_{g,\ell}$ is given by   
    \begin{align}
        f_{g,\ell} &= \frac{\partial \phi}{\partial \tau} \delta_{0\ell} + \frac{k}{3} \frac{\epsilon}{q} \psi \delta_{1\ell} \, ,
    \end{align}
where $\phi$ and $\psi$ are the two gravitational potentials, while $\delta_{0\ell}$ and $\delta_{1\ell}$ are Kronecker deltas selecting the monopole and dipole contributions, respectively.\footnote{This notation was introduced in Ref.~\cite{Perez2026towards} and can be straightforwardly adapted to the synchronous gauge.}

The right-hand side of Eq.~\eqref{Eq: BHE_CN} depends on the effective coupling $G_{\rm eff}=g^2/m_\varphi^2$, where $g$ is the fundamental coupling, $m_\varphi$ the mediator mass, and on the dimensionless function $H_{\nu,\ell}(\epsilon)$, defined as~\cite{Kreisch2020, Perez2026towards},
    \begin{equation}
        H_{\nu,\ell}(\epsilon) = \frac{1}{f_\nu^{(0)}} \left(  \frac{T_{\nu,0}}{\epsilon}\right) \left[ A (q_1) + B_\ell (q_1) - 2 D_\ell (q_1)\right] \, , \label{eq:H-function}
    \end{equation}
for the $\nu\nu\to\nu\nu$ scattering contribution between neutrinos in the same mass eigenstate. Here, $q_1=p_1/T_\nu= q/T_{\nu,0}$, is the scaled momentum of the reference neutrino, where $T_{\nu}$ and $T_{\nu,0}$ denote the neutrino temperature and its present-day value, respectively.

As shown in Ref.~\cite{Perez2026towards}, neutrino masses $m_\nu \lesssim 0.06 \,{\rm eV}$ lead to only small deviations for $T_\nu\lesssim 1 \,{\rm eV}$ in the heavy mediator limit for $m{_\varphi}\gtrsim 1 \,{\rm keV}$, 
considering the simplest neutrino-scalar interaction. Therefore, one can use the approximation $T_{\nu,0}/\epsilon\simeq 1/q_1$, such that $H_{\nu,\ell}(\epsilon) \simeq H_{\nu,\ell}(q_1)$. 
In the simplified model described by the BHEs in Eq.~\eqref{Eq: BHE_CN}, we only consider one process, $\nu_i \nu_i \to \nu_i \nu_i$ ($i=1,2,3$), under which the first-order collision term is the same for all reference neutrinos and therefore $\nu_{i, \ell} = \nu_{\ell}$ and is described by Eq.~\eqref{eq:H-function}.
Even though diagonal coupling allows processes such as $\nu_1 \nu_1 \to \nu_2 \nu_2$, we expect these contributions to be proportional to the contribution considered here and affect our subsequent analysis only through a simple rescaling of $G_{\rm eff}$~\cite{Perez2026towards}.
\section{\label{sec:data} Data and Methodology}

In this section, we briefly describe the datasets and 
combinations, as well as the analysis setup employed throughout this work.

\subsection{Datasets}

\begin{enumerate}

    \item \textbf{DESI:}
    We employ three different types of data.
    
    \textit{DESI DR1 FS+BAO:} 
    We use the DESI DR1 full-shape (FS) 
    measurements of the redshift-space galaxy and quasar power spectrum, combined with the DR1 post-reconstruction BAO measurements through the public joint FS+BAO clustering products.\footnote{\url{https://data.desi.lbl.gov/doc/releases/dr1/vac/full-shape-bao-clustering/}} The published joint covariance accounts for FS--BAO correlations~\cite{adame2024desiFS,DESI:2024jis}.
    The samples comprise the Bright Galaxy Sample (BGS), luminous red galaxies (LRGs), emission line galaxies (ELGs), and quasars (QSOs), divided into six tracer--redshift bins spanning $0.1<z<2.1$ and covering an effective sky area around 7,500 deg$^2$.

    \textit{DESI BAO:} 
    We use BAO distance measurements from DESI DR1~\cite{adame2024desiBAO} or DR2~\cite{Abdul2025desi}.\footnote{We pair DR1 FS with DR1 BAO to retain their published cross-covariance. Since the cross-covariance between DR1 FS and DR2 BAO is unavailable, DR2 BAO enters only the BAO+CMB-SPA combination. For joint analyses using alternative treatments of statistical overlap, see Refs.~\cite{Chudaykin:2025aux, Forero-Sanchez:2026bff, Forero-Sanchez:2026zfb}.} Depending on the tracer and release, the measurements constrain $D_M(z)/r_d$ and $D_H(z)/r_d$, or $D_V(z)/r_d$, where $D_M$ is the transverse comoving distance, $D_H=c/H(z)$, $D_V=[zD_M^2D_H]^{1/3}$, and $r_d$ is the sound horizon at the baryon drag epoch. 
    Both releases include galaxy and quasar clustering and Ly-$\alpha$ forest BAO at $z_{\rm eff}=2.33$.

    \textit{DESI Ly-$\alpha$ DR1:} 
    We additionally use cosmological constraints derived from DESI DR1 measurements of the one-dimensional Ly-$\alpha$ forest flux power spectrum~\cite{Chaves-Montero_2026_DESIlya}.\footnote{\gitlinklyalpha} These measurements extend the redshift coverage up to $z\approx4$ and probe smaller scales than galaxy clustering observations, reaching $k\sim4\ h\ {\rm Mpc}^{-1}$.

    \item  \textbf{CMB-SPA:}
    For our primary CMB dataset, we adopt the CMB-SPA data combination introduced in \cite{SPT_2026_CMBSPA}, which combines measurements from SPT-3G~\cite{SPT_2026_CMBSPA}, Planck~\cite{Planck:2019nip}, and ACT DR6~\cite{Louis_2025_ACT_DR6}. We include the Planck low-$\ell$ temperature and polarization data, with the latter providing a direct constraint on the optical depth to reionization, $\tau_{\rm reio}$~\cite{deBelsunce:2021mec}, instead of adopting a Gaussian prior as in~\cite{SPT_2026_CMBSPA}. Together, the Planck, ACT DR6, and SPT-3G measurements provide broad multipole coverage of the TT, TE, and EE power spectra, combining large-scale CMB information with precise measurements at intermediate and small angular scales.

    \item \textbf{External priors:}
    For fits without a CMB likelihood, we follow the DESI analysis~\cite{DESI:2024jis} and impose a Gaussian Big Bang nucleosynthesis (BBN) prior on the physical baryon density, $\omega_b=0.02218\pm0.00055$~\cite{Schoneberg_2024}. We also apply the weak Gaussian prior $n_s=0.9649\pm 0.042$, denoted by $n_{s10}$. Its mean is taken from \textit{Planck 2018}~\cite{Planck:2018vyg}, and its standard deviation is enlarged by a factor of ten. Both priors are listed in Table~\ref{table:priors} and are removed when a CMB likelihood is included.

\end{enumerate}
We investigate four main dataset combinations. Two of these combine DESI LSS data with minimal priors on $\omega_b$ and $n_s$, as described above. 
These combinations are DESI DR1 (FS+BAO)+BBN +$n_{s10}$ and DR1 (FS+BAO+Ly-$\alpha$)+BBN+$n_{s10}$. The other two combinations include CMB data: DESI BAO DR2+CMB-SPA and (FS+BAO) DR1+CMB-SPA.
We also compare our results with constraints reported in previous studies based on earlier galaxy survey and CMB datasets.

\subsection{Analysis setup}

\begin{table*}
\begingroup
\renewcommand{\arraystretch}{1.15}
\begin{ruledtabular}
\begin{tabular}{lccl}
Parameter & Default & Prior & Comments \\
\hline
\multicolumn{4}{l}{\textbf{Cosmological}} \\
$\log_{10}(G_{\rm eff}\,{\rm MeV}^{2})$ for MI$_\nu$
& --- & $\mathcal{U}(-5.5,-2)$ & --- \\
$\log_{10}(G_{\rm eff}\,{\rm MeV}^{2})$ for SI$_\nu$
& --- & $\mathcal{U}(-2,0.5)$ & --- \\
$N_{\rm eff}$
& $3.046$ & $\mathcal{U}(2.0,5.0)$ & --- \\
$\sumnu\;[{\rm eV}]$
& $0.06$ & $\mathcal{U}(0,2)$ & --- \\
$H_0\;[\kmsMpc]$
& --- & $\mathcal{U}(20,100)$ & --- \\
$\ocdm$
& --- & $\mathcal{U}(0.01,0.99)$ & --- \\
$\ob$
& --- & $\mathcal{N}(0.02218,0.00055^2)$
& BBN prior \\
$\ln(10^{10}A_s)$ 
& --- & $\mathcal{U}(2.0,4.0)$ &  --- \\
$n_s$
& --- & $\mathcal{N}(0.9649,0.042^2)$
& $n_{s10}$: \textit{Planck} $10\sigma$ \\
\hline
\multicolumn{4}{l}{\textbf{Cosmological (with CMB)}} \\
$100\theta_s$
& --- & $\mathcal{U}(0.5,10)$ & sampled instead of $H_0$ \\
$\tau_{\rm reio}$
& $0.0544$ & $\mathcal{U}(0.01,0.8)$ & varied with CMB; fixed otherwise \\
$\ob$
& --- & $\mathcal{U}(0.005, 0.1)$
& no BBN prior \\
$n_s$
& --- & $\mathcal{U}(0.8, 1.2)$
& no $n_{s10}$ prior \\
\hline
\multicolumn{4}{l}{\textbf{FS nuisances}} \\
$b_1\sigma_8$
& & $\mathcal{U}(0,3)$ & each $z$-bin \\
$b_2\sigma_8^2$
& & $\mathcal{N}(0,5^2)$ & each $z$-bin \\
$b_s\sigma_8^2$
& & $\mathcal{N}(0,5^2)$ & each $z$-bin \\
$\alpha_0$
& & $\mathcal{N}(0,12.5^2)$ & each $z$-bin$^{*}$ \\
$\alpha_2$
& & $\mathcal{N}(0,12.5^2)$ &  each $z$-bin$^{*}$ \\
$\alpha^{\rm shot}_0$
& & $\mathcal{N}(0,2^2)$ &  each $z$-bin$^{*}$ \\
$\alpha^{\rm shot}_2$
& & $\mathcal{N}(0,5^2)$ &  each $z$-bin$^{*}$ \\
\end{tabular}
\end{ruledtabular}
\endgroup
\caption{Parameters and priors adopted in our analyses. The default column gives the fixed value used when the corresponding cosmological parameter is not varied. We use $\mathcal{U}({\rm min},{\rm max})$ to denote a uniform prior over the given range, and $\mathcal{N}(\mu,\sigma^2)$ to denote a Gaussian distribution with mean $\mu$ and standard deviation $\sigma$. The asterisk denotes the terms that are analytically marginalized over, following~\cite{Noriega:2024eyu}.
}
\label{table:priors}
\end{table*}

To explore the NSI model described by Eqs.~\eqref{Eq: BHE_CN}, we compute linear perturbations using the public modified \textsc{CLASS} implementation of Ref.~\cite{Camarena2023confronting}.\footnote{\gitlinkcato} We then feed the resulting linear power spectrum into \textsc{Folps}\footnote{\gitlinkfolps} to compute the galaxy redshift-space power spectrum at one-loop order~\cite{Folps_Noriega_2022, DESI:2026haa}. Its $\texttt{fk}$-kernels go beyond the Einstein--de Sitter approximation by retaining the dominant contributions associated with the scale-dependent growth induced by massive neutrinos~\cite{Folps_Noriega_2022}. Since neutrino self-interactions become negligible while perturbations on the scales considered here are still linear, we neglect additional NSI corrections to late-time nonlinear evolution~\cite{Camarena2023confronting, He2024self, Camarena2025strong}.

When fitting the different dataset combinations, we sample over the cosmological parameters ${H_0, \, \omega_{\rm cdm}, \,\omega_{\rm b},\, \ln(10^{10}A_s)}$ and $n_s$, where $H_0$ is the Hubble constant, $\omega_{\rm cdm}=\Omega_{\rm cdm}h^2$ and $\omega_{\rm b}=\Omega_{\rm b}h^2$ are the present-day physical densities of cold dark matter and baryons, respectively, $A_s$ and $n_s$ are the amplitude and spectral index of the primordial scalar perturbations. For analyses including CMB data, we additionally sample the optical depth to reionization, $\tau_{\rm reio}$, and replace $H_0$ with $100\theta_s$ as a sampled parameter, treating $H_0$ as a derived quantity.
For the neutrino sector, we assume three degenerate massive neutrinos (see e.g.,~\cite{Hamann2012measuring, Brinckmann2019promising}). 
In the baseline scenario, we fix $\sum m_\nu=0.06 \,{\rm eV}$ and $N_{\rm eff}=3.046$. However, we also study the impact of allowing $\sumnu$ and $N_{\rm eff}$ to vary. When NSI is considered, the parameter space is extended to include $\log_{10}(G_{\rm eff}\,{\rm MeV}^{2})$.
The adopted parameter priors are summarized in Table~\ref{table:priors}, where we also provide comments on specific choices.

In addition to the cosmological parameters, we include a set of nuisance parameters for each tracer and redshift bin in the full-shape analyses. These comprise the Eulerian galaxy bias parameters $b_1$, $b_2$, and $b_s$, sampled through the combinations $b_1\sigma_8(z)$, $b_2\sigma_8^2(z)$, and $b_s\sigma_8^2(z)$. This amplitude-scaled parametrization is motivated by the DESI analysis~\cite{adame2024desiFS} and aims to mitigate prior-volume effects. We also include the EFT counterterms $\alpha_{0,2}$ and the stochastic coefficients $\alpha^{\rm shot}_{0,2}$, which enter the power-spectrum model linearly and are marginalized over analytically, with their adopted priors, following Appendix~A of Ref.~\cite{Noriega:2024eyu}. 

For the data analysis, we perform profile likelihood analyses using the \textsc{bobyqa} minimizer~\cite{Powell2009_BOBYQA} through \textsc{cobaya}~\cite{Torrado_2021_cobaya}, which we also use for Markov chain Monte Carlo (MCMC) sampling. MCMC convergence is assessed using the Gelman--Rubin criterion $R-1< {\cal O}(10^{-2})$~\cite{Gelman:1992zz}. We use \textsc{getdist}~\cite{Getdist_Lewis_2025} to obtain marginalized posterior distributions and parameter constraints.

\section{\label{sec:results_lss} Constraints from DESI DR1
}

In this section, we investigate CMB-independent constraints on the neutrino self-interaction from the latest publicly available datasets DESI DR1 data supplemented by BBN and $n_{s10}$ priors.

\subsection{Profile likelihood}

\begin{figure}
    \centering
    \includegraphics[width=1\linewidth]{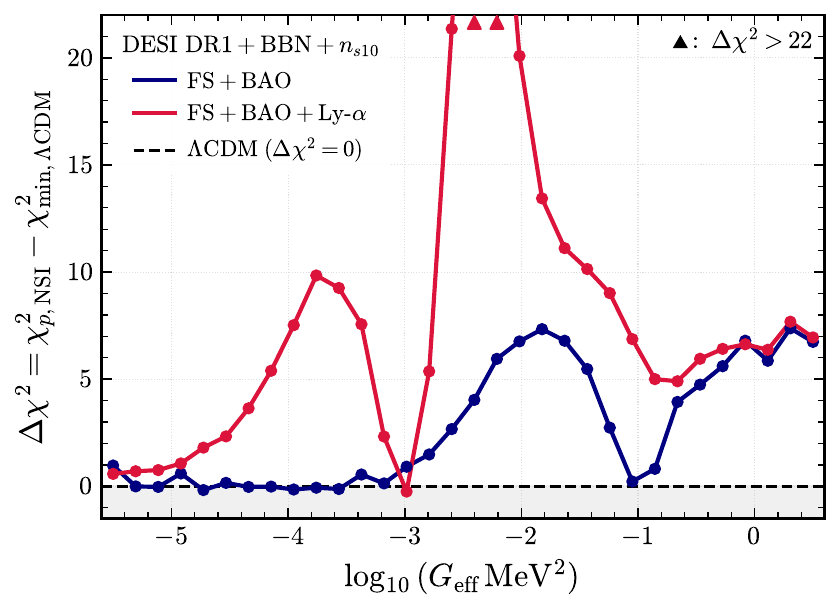}
    \caption{Profile likelihood (normalized to $\Lambda$CDM) of the NSI parameter $\log_{10}(G_{\rm eff\,} {\rm MeV^{2}})$, obtained from the analyses of DESI DR1 (FS+BAO)+BBN+$n_{s10}$ (blue) and DESI DR1 (FS+BAO+Ly-$\alpha$)+BBN+$n_{s10}$ (red). Here, we kept the total neutrino mass and effective number of relativistic species fixed at $\sum m_\nu=0.06\,\mathrm{eV}$ and $N_{\rm eff} = 3.046$.}
    \label{fig:profile}
\end{figure}

Our first objective is to search for distinct maxima in the likelihood as a function of $G_{\rm eff}$, or equivalently, distinct minima in $\chi^2$. 
We therefore compute a profile likelihood by fixing $\log_{10}(G_{\rm eff}\,\mathrm{MeV}^{2})$ at 32 uniformly spaced values
between $-5.5$ and $0.5$. At each grid point, we use the \textsc{bobyqa} minimizer through \textsc{cobaya} to minimize $\chi^2$ over the remaining cosmological and nuisance parameters. We perform three separate minimization runs with different initial parameter values, and keep the lowest $\chi^2$ obtained. This reduces sensitivity to the choice of starting point and to convergence to local minima.
We express the resulting profiles relative to the corresponding minimum value of $\chi^2$ for $\Lambda$CDM, such that distinct preferred regions of $G_{\rm eff}$ appear as separate local minima in the profile.

Figure~\ref{fig:profile} shows these profiles for DESI DR1 (FS+BAO)+BBN+$n_{s10}$ and for the same combination including Ly-$\alpha$. For DESI DR1 (FS+BAO)+BBN+$n_{s10}$, two distinct preferred regions appear: a nearly flat moderate-interaction regime and a strongly interacting local minimum at $\log_{10}(G_{\rm eff}\,\mathrm{MeV}^{2})\simeq-1$. Neither region provides an appreciable improvement over $\Lambda$CDM. This result is consistent with the baseline full-shape analysis of Ref.~\cite{Poudou:2025qcx}. That paper also showed that the relative preference for the ${\rm SI}_\nu$ solution depends on the priors imposed on the EFT nuisance parameters. In particular, adopting least informative priors
recovers the ${\rm SI}_\nu$ preference over $\Lambda$CDM, as reported in \cite{Camarena2023confronting}.\footnote{We expect the effect of EFT priors to be negligible when full-shape data are combined with CMB.
}

Adding Ly-$\alpha$ changes the profile substantially. Now we observe three candidate modes: a weakly interacting branch connected to $\Lambda$CDM, a narrow minimum near $\log_{10}(G_{\rm eff}\,\mathrm{MeV}^{2})\simeq-3$, and a strongly interacting minimum near $\log_{10}(G_{\rm eff}\,\mathrm{MeV}^{2})\simeq-1$.
However, we noticed that the strong mode is disfavored by the inclusion of the Ly-$\alpha$ data with $\Delta \chi^2 \sim 5$; thus, we expect this to be penalized in the MCMC and mode comparison analyses. 
The mode appearing at $\log_{10}(G_{\rm eff\,} {\rm MeV^{2}})\simeq -3$ identifies a distinct solution at finite interaction strength. However, its minimum $\chi^2$
is comparable to that of the weakly interacting branch and
provides no appreciable improvement over $\Lambda$CDM.

Overall, the DESI DR1 profiles provide further evidence for a multimodal likelihood structure, consistent with previous CMB-independent analyses of galaxy clustering. The inclusion of Ly-$\alpha$ reveals an additional, narrow
minimum near $\log_{10}(G_{\rm eff}\,\mathrm{MeV}^{2})\simeq-3$ calling for further investigation.

\subsection{MCMC analysis}

\begin{figure*}
    \centering
    \resizebox{\linewidth}{!}{
        \includegraphics[height=0.5\linewidth]{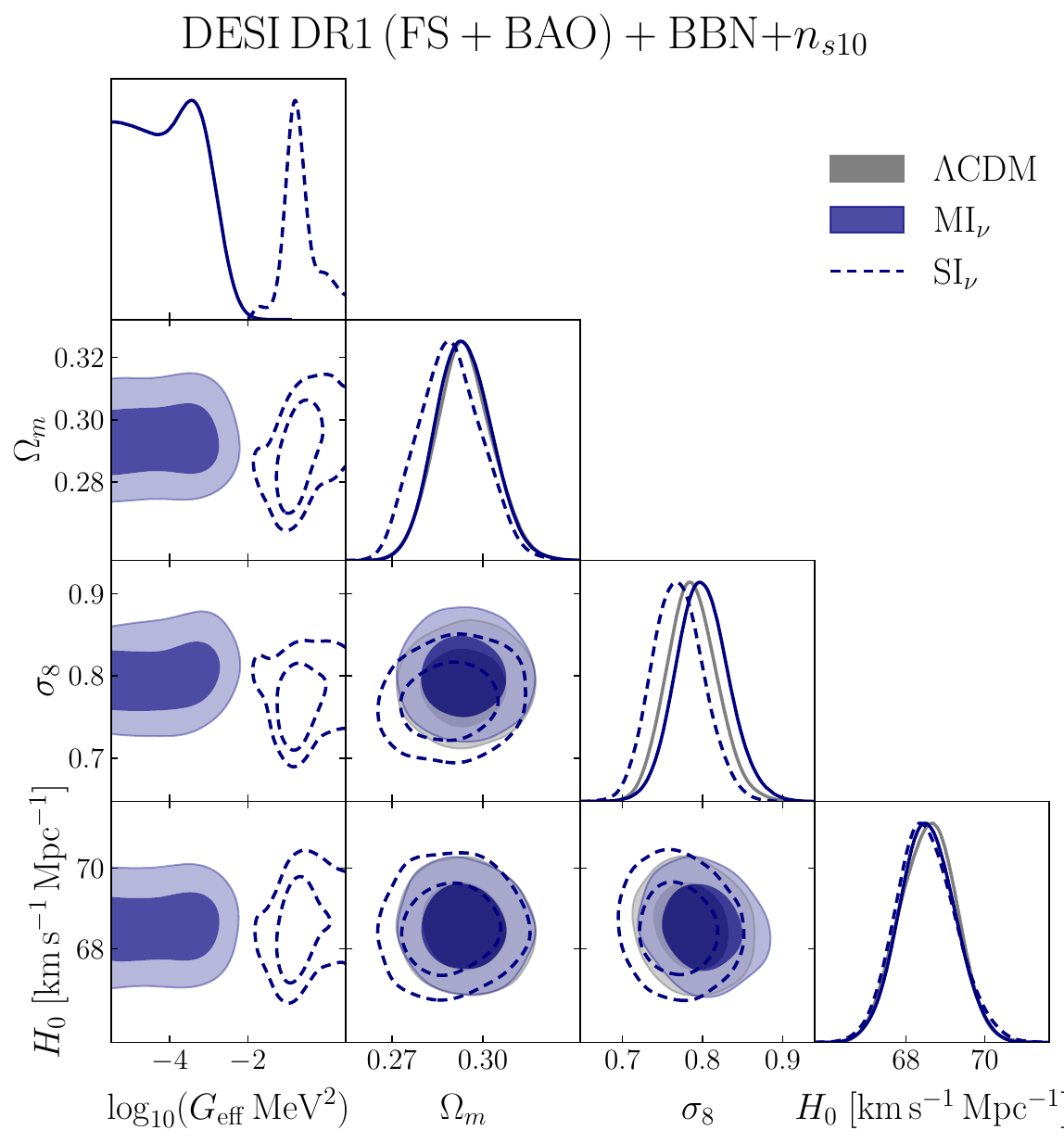}
        \includegraphics[height=0.5\linewidth]{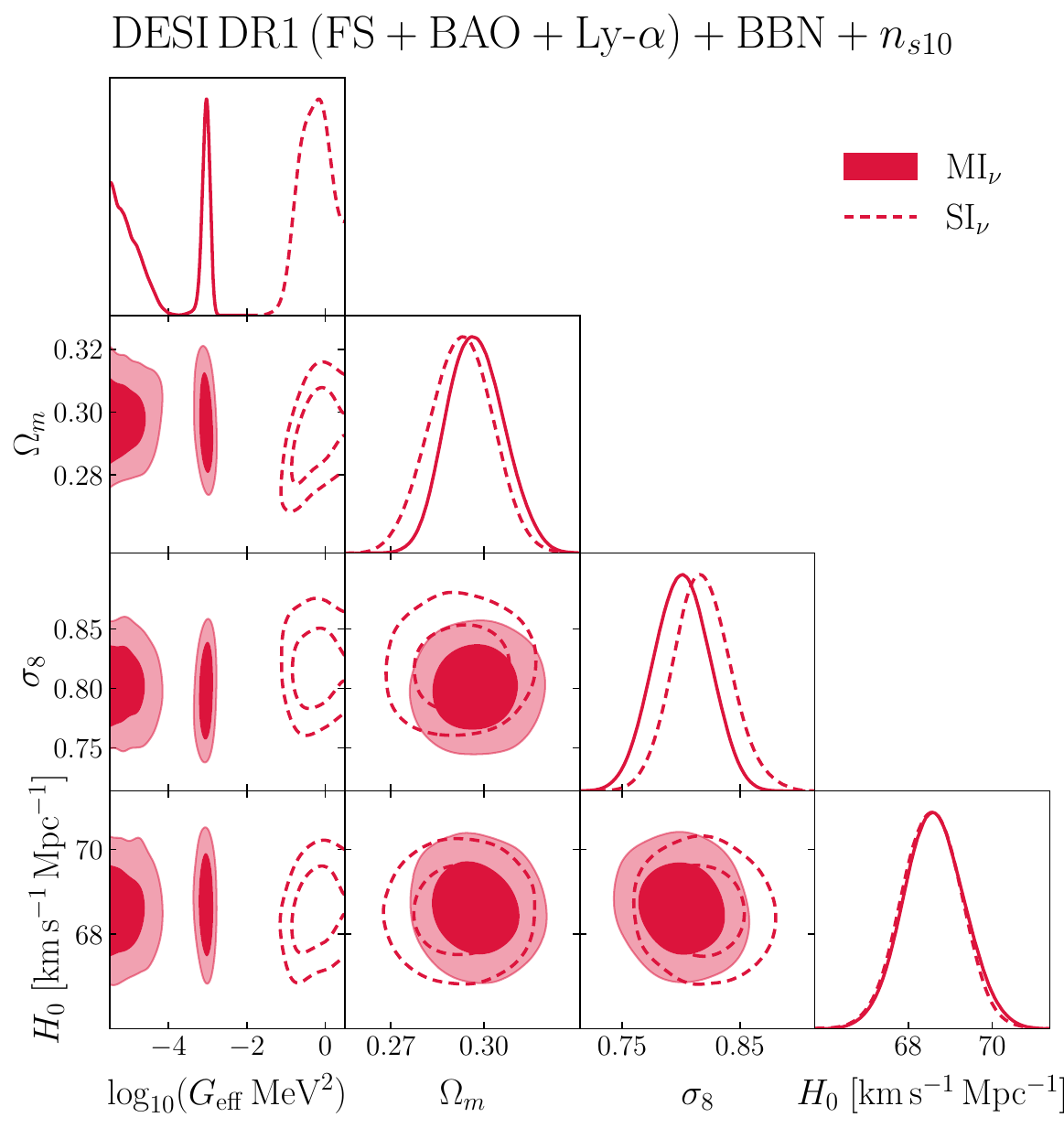}
    }
    \caption{Posterior distributions at 68\% and 95\% CLs on the cosmological parameters of the NSI model with fixed neutrino mass, considering DESI DR1 (FS+BAO)+BBN+$n_{s10}$ (left), and including Ly-$\alpha$ data (right). For comparison, we also show the $\Lambda$CDM bounds (gray) on the left. The corresponding parameter constraints are reported in Table~\ref{tab:cosmo}. The full contour plot for the FS+BAO case, including a comparison with free neutrino mass, is shown in Fig.~\ref{fig:appen_fs_mass_nu}.}   
    \label{fig:results_lss}
\end{figure*}

We next derive parameter constraints from DESI DR1 (FS+BAO)+BBN+$n_{s10}$. To address the bimodality of the posterior distribution of $G_{\rm eff}$, we follow previous studies (e.g.,~\cite{Choudhury2021, Calabrese2025atacama, Poudou:2025qcx}) and perform separate MCMC analyses of the moderately interacting and strongly interacting modes, adopting distinct prior ranges for $G_{\rm eff}$. Guided by our profile-likelihood results, we fix the boundary between the two modes at $\log_{10}(G_{\rm eff} \,\mathrm{MeV}^{2})=-2$.

We show our results in Table \ref{tab:cosmo} and in the left panel of Fig. \ref{fig:results_lss}. 
First, notice that the posterior distribution $\log_{10}(G_{\rm eff\,} {\rm MeV^{2}})$ is not normalized over the full range from $-5.5$ to $0.5$ due to the mode separation. Therefore, the peak heights should not be interpreted as indicating the relative significance of the modes; a comparison using Bayesian evidence is discussed in Section~\ref{sec:bayesian}.
Next, we discuss the highlights of these bounds: In general, the results of the MI$_\nu$ mode are in good agreement with a non-NSI $\Lambda$CDM model, as expected, no significant deviations are found in any of the cosmological parameters. 
When $\sum m_\nu$ is allowed to vary freely, the constraints on $\log_{10}(G_{\rm eff\,} {\rm MeV^{2}})$ and the other cosmological parameters remain broadly similar, as shown in Table~\ref{tab:cosmo} and in Fig.~\ref{fig:appen_fs_mass_nu}. The main exception is $\Omega_m$, which shifts to slightly higher values and becomes less tightly constrained due to its well-known degeneracy with $\sum m_\nu$~\cite{adame2024desiFS}.
Our bound of $\sum m_\nu<0.396\, {\rm eV}$ at $95\%$ confidence level (CL) is similar to the limit of $\sum m_\nu<0.409\, {\rm eV}$ at $95\%$ CL reported by the DESI Collaboration within flat $\Lambda$CDM, using full-shape and BAO data with a BBN prior and the $10\sigma$ prior on $n_s$~\cite{adame2024desiFS}.

We find that the SI$_\nu$ mode favors slightly lower $\sigma_8$ and $\Omega_m$ than $\Lambda$CDM and MI$_\nu$, consistent with the BOSS results of~\cite{Poudou:2025qcx, Camarena2023confronting}. Fig.~\ref{fig:appen_fs_mass_nu} also shows small downward shifts in $A_s$ and $n_s$ relative to MI$_\nu$, in agreement with~\cite{Camarena2023confronting}. Both interaction regimes yield similar $H_0$ values consistent with $\Lambda$CDM, even when $\sumnu$ is allowed to vary. Notably, the later onset of neutrino free streaming in SI$_\nu$ relaxes the 95\% CL upper bound on $\sumnu$ by approximately $29\%$, from $0.396\,\mathrm{eV}$ in MI$_\nu$ to $0.509\,\mathrm{eV}$.

\subsection{\label{sec:lyalpha} Inclusion of Ly-\texorpdfstring{$\alpha$}{alpha} into the MCMC analysis}

\begin{figure}
    \centering
    \includegraphics[width=1\linewidth]{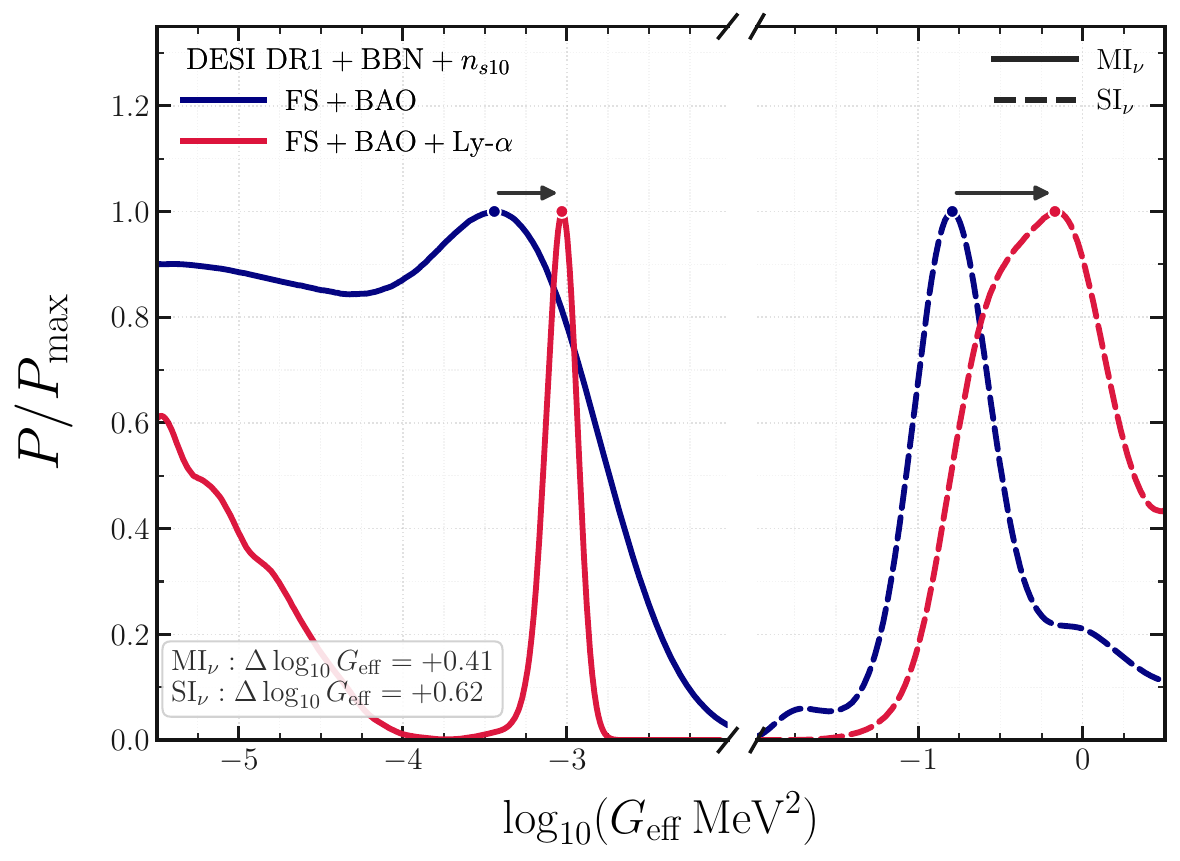}
    \caption{Comparison of the one-dimensional marginalized posteriors of $\log_{10}(G_{\rm eff\,} {\rm MeV^{2}})$ obtained from DESI DR1 (FS+BAO)+BBN+$n_{s10}$ (blue) and from the same dataset including Ly-$\alpha$ data (red). The latter shifts the posteriors toward larger values of the neutrino self-interaction coupling. }
    \label{fig:1d_posterior_LSS}
\end{figure}

\begin{table*}
\caption{Parameter constraints from DESI datasets combined with a BBN prior on $\omega_b$ and a weak Gaussian prior on $n_s$ with a standard deviation ten times the Planck uncertainty. Parameter uncertainties are reported at 68\% CL and upper bounds at 95\% CL. With Ly-$\alpha$, the posterior within the MI$_\nu$ regime has two modes: weakly interacting (WI$_\nu$) and intermediate interacting (II$_\nu$). We report an upper bound for WI$_\nu$ and a measurement for II$_\nu$, separating the modes at $\log_{10}(G_{\rm eff}\,{\rm MeV}^{2})=-3.735$ after the MCMC runs.
}
\label{tab:cosmo}
\begingroup
\renewcommand{\arraystretch}{1.25}
\setlength{\tabcolsep}{4pt}
\begin{ruledtabular}
\begin{tabular}{lccccc}
Model/Dataset
&
$\log_{10}(G_{\rm eff}\,{\rm MeV}^{2})$ &
$H_0$ [$\mathrm{km\,s^{-1}\,Mpc^{-1}}$] &
$\Omega_m$ &
$\sigma_8$ &
$\sum m_\nu$ [$\mathrm{eV}$] \\[0.5ex]
\colrule
\multicolumn{6}{l}{\rule{0pt}{2.8ex}$\bm{\Lambda}$\textbf{CDM}}\\[0.5ex]
\hspace{1em}\shortstack[l]{DESI DR1 (FS+BAO)\\+BBN+$n_{s10}$} & ---& $68.56\pm 0.71$ & $0.2939\pm 0.0093$ & $0.787\pm 0.031$ & ---\\[0.5ex]
\colrule
\multicolumn{6}{l}{\rule{0pt}{2.8ex}$\bm{\Lambda}$\textbf{CDM}+$\bm{G_{\rm eff}}$}\\[0.5ex]
\hspace{1em}\shortstack[l]{DESI DR1 (FS+BAO)\\+BBN+$n_{s10}$} & & & & & \\[0.5ex]
\hspace{1.5em}MI$_\nu$ & $< -2.70$ & $68.54\pm 0.70$ & $0.2938^{+0.0087}_{-0.0097}$ & $0.800\pm 0.033$ & --- \\[0.4ex]
\hspace{1.5em}SI$_\nu$ & $-0.67^{+0.27}_{-0.47}$ & $68.51^{+0.70}_{-0.80}$ & $0.289^{+0.010}_{-0.011}$ & $0.770^{+0.030}_{-0.034}$ & --- \\[0.9ex]
\hspace{1em}\shortstack[l]{DESI DR1 (FS+BAO+Ly-$\alpha$) \\+BBN+$n_{s10}$} & & & & & \\[0.5ex]
\hspace{1.5em}MI$_\nu$ &
\begin{tabular}{@{}c@{\hspace{1em}}c@{}}
$<-4.39$ &
$-3.04^{+0.11}_{-0.078}$
\end{tabular}
&
$68.62\pm 0.72$ &
$0.2973^{+0.0086}_{-0.0097}$ &
$0.801\pm 0.023$ &
--- \\[0.4ex]
\hspace{1.5em}SI$_\nu$ & $-0.26\pm 0.39$ & $68.55\pm 0.71$ & $0.293\pm 0.010$ & $0.818\pm 0.024$ & --- \\[0.9ex]
\colrule
\multicolumn{6}{l}{\rule{0pt}{2.8ex}$\bm{\Lambda}$\textbf{CDM}+$\bm{G_{\rm eff}}$+$\bm{\sum m_\nu}$}\\[0.5ex]
\hspace{1em}\shortstack[l]{DESI DR1 (FS+BAO)\\+BBN+$n_{s10}$} & & & & & \\[0.5ex]
\hspace{1.5em}MI$_\nu$ & $< -2.70$ & $68.49\pm 0.71$ & $0.299^{+0.010}_{-0.013}$ & $0.799^{+0.030}_{-0.035}$ & $< 0.396$ \\[0.4ex]
\hspace{1.5em}SI$_\nu$ & $-0.76^{+0.22}_{-0.39}$ & $68.34\pm 0.74$ & $0.296^{+0.012}_{-0.014}$ & $0.775^{+0.030}_{-0.033}$ & $< 0.509$ \\[0.9ex]
\end{tabular}
\end{ruledtabular}
\endgroup
\end{table*}

Motivated by the profile-likelihood results discussed above, we extend our LSS analysis to include the Ly-$\alpha$. This extension is particularly relevant given recent studies showing that Ly-$\alpha$ measurements can substantially affect the inferred constraints on $G_{\rm eff}$, with the results depending on the adopted Ly-$\alpha$ likelihood~\cite{He2024self, He:2025jwp}.
We present the resulting parameter constraints in Table~\ref{tab:cosmo} and the right panel of Fig.~\ref{fig:results_lss}. The 1D $\log_{10}(G_{\rm eff\,} {\rm MeV^{2}})$ posterior is shown in Fig.~\ref{fig:1d_posterior_LSS} for further details.

We now turn to the moderately interacting regime.
As anticipated from the profile likelihood analysis, including Ly-$\alpha$ data leads to a bimodal marginalized posterior for $\log_{10}(G_{\rm eff}\,{\rm MeV}^{2})$ within this regime.
To obtain separate NSI constraints for the two modes identified within this regime, we split samples at $\log_{10}(G_{\rm eff}\,{\rm MeV}^{2})=-3.735$ after the MCMC runs were completed.\footnote{This value corresponds to the local minimum of the marginalized posterior density between the two peaks, providing a natural boundary for separating the modes.}
We propose to name the left side as the weakly interacting (WI$_\nu$) mode, and the right side as the intermediate interacting (II$_\nu$) mode.
In the WI$_\nu$ mode, we find an upper bound at $\log_{10}(G_{\rm eff}\,{\rm MeV}^{2})<-4.39$ (95\% CL), similar to the bound of $\log_{10}(G_{\rm eff}\,{\rm MeV}^{2})<-4.40$ (95\% CL) reported in Ref.~\cite{He:2025jwp}, obtained by combining Planck CMB data with an EFT-based likelihood for the eBOSS Ly-$\alpha$ 1D power spectrum.
Regarding the II$_\nu$ mode, we find the following constraint $\log_{10}(G_{\rm eff\,} {\rm MeV^{2}})=-3.04^{+0.11}_{-0.078}$ (68\% CL), which represents a very sharp Gaussian-like distribution whose central value is located at more than 8$\sigma$ from the adopted lower boundary of the II$_\nu$ mode.

Besides, notice that both mode peaks are shifted to larger $G_{\rm eff}$-values compared to the case without Ly-$\alpha$  
(see Fig.~\ref{fig:1d_posterior_LSS}).
As for the other cosmological parameters, the most noticeable change is in $\sigma_8$: its central value remains essentially unchanged, while its uncertainty reduces by 30\% with respect to data without Ly-$\alpha$.

For the SI$_{\nu}$ mode, Table~\ref{tab:cosmo} shows that the NSI parameter, $G_{\rm eff}$, moves toward larger values compared with the analysis without Ly-$\alpha$ data.
It also leads to a larger $\sigma_8$, while $\Omega_m$ increases slightly and $H_0$ remains nearly unchanged. Comparing instead with the moderately interacting mode, the SI$_{\nu}$ mode favors a slightly larger $\sigma_8$ and a lower $\Omega_m$, which have competing effects on $\sigma_8$-tension.

Finally, Ly-$\alpha$ forest measurements probe scales extending into the nonlinear regime, and their cosmological interpretation requires accurate modeling of nonlinear evolution and intergalactic-medium physics~\cite{Chaves-Montero_2026_DESIlya}. For example, Ref.~\cite{He:2025jwp} finds that, in combination with Planck data, the PRIYA-based likelihood allows larger values of $G_{\rm eff}$ and retains a secondary strongly interacting mode that is strongly disfavored by the EFT-based likelihood. This comparison illustrates the sensitivity of the inferred posterior to the Ly-$\alpha$ likelihood and its modeling assumptions. It remains unclear whether the bimodality found within the MI$_\nu$ region in our analysis would persist under alternative treatments of nonlinear evolution and intergalactic-medium physics.

\section{\label{sec:results_cmb} Constraints from DESI and CMB
}
In this section, we focus on the results from the joint DESI and CMB data.
The inclusion of CMB data is important since the NSI anomalous bimodality appeared first using WMAP data~\cite{CyrRacine2014}, then reappeared with Planck~\cite{Lancaster2017, Kreisch2020, Park2019} and ACT data \cite{Kreisch:2022zxp, Calabrese2025atacama}.
To our knowledge, this work provides the first constraints on the NSI model using the full CMB-SPA likelihood.
The first dataset we use consists of  DESI DR2 BAO+CMB-SPA, and after DESI DR1 (FS+BAO)+CMB-SPA.

\subsection{DESI DR2 BAO and CMB-SPA joint analysis}

\begin{figure*}
    \centering
    \includegraphics[width=1\linewidth]{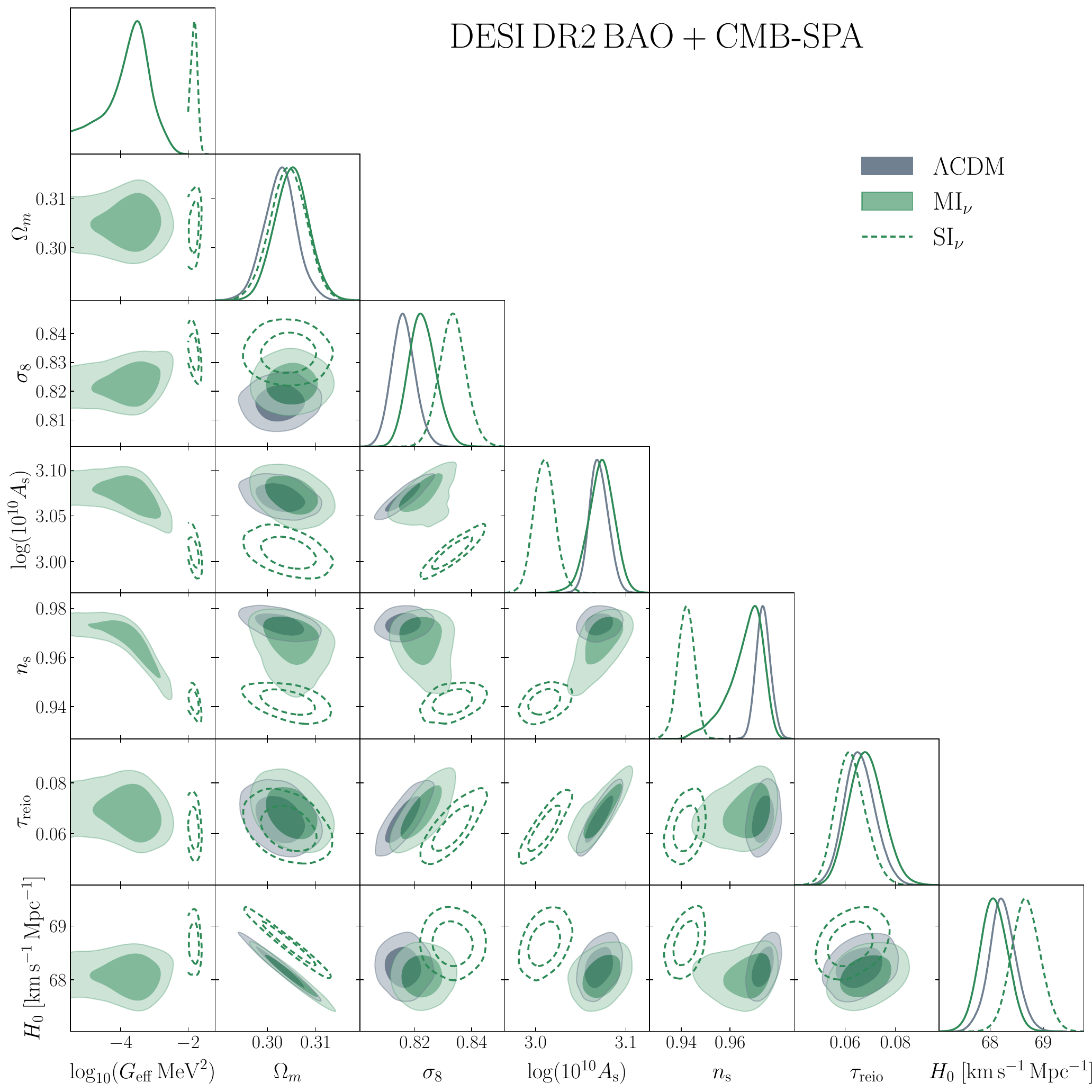}
    \caption{Posterior distributions at 68\% and 95\% CLs on the cosmological parameters of the NSI model with fixed neutrino mass, considering DESI DR2 BAO+CMB-SPA. For comparison, we also show the $\Lambda$CDM bounds (gray). The corresponding parameter constraints are reported in Table~\ref{tab:cosmo_cmb}. The full contour plot, including a comparison with free $\sumnu$ and $N_{\rm eff}$, is shown in Fig.~\ref{fig:appen_desi_cmb_neff_mnu}.}
    \label{fig:cmb_bao}
\end{figure*}

\begin{figure*}
    \centering
    \resizebox{\linewidth}{!}{
        \includegraphics[height=0.5\linewidth]{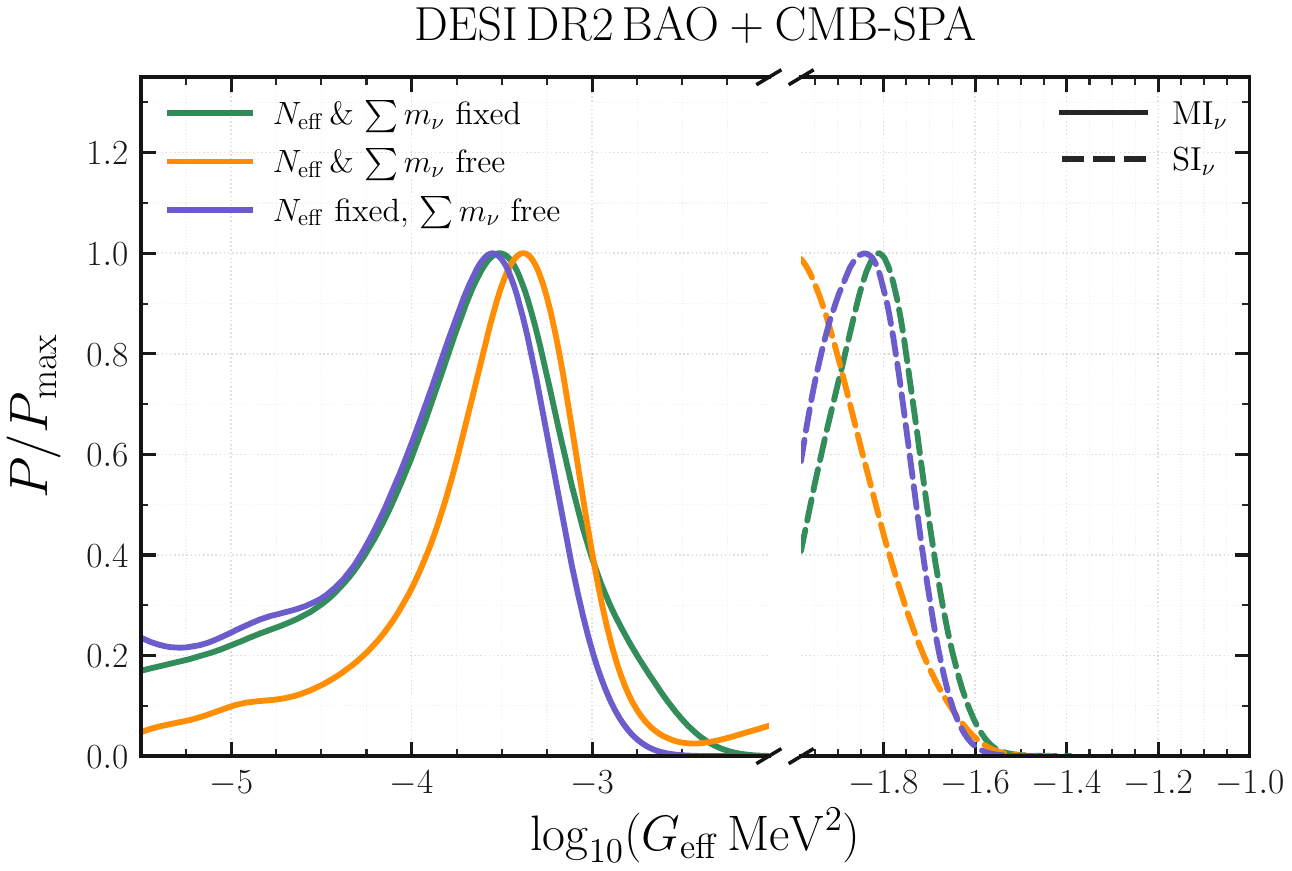}
        \includegraphics[height=0.5\linewidth]{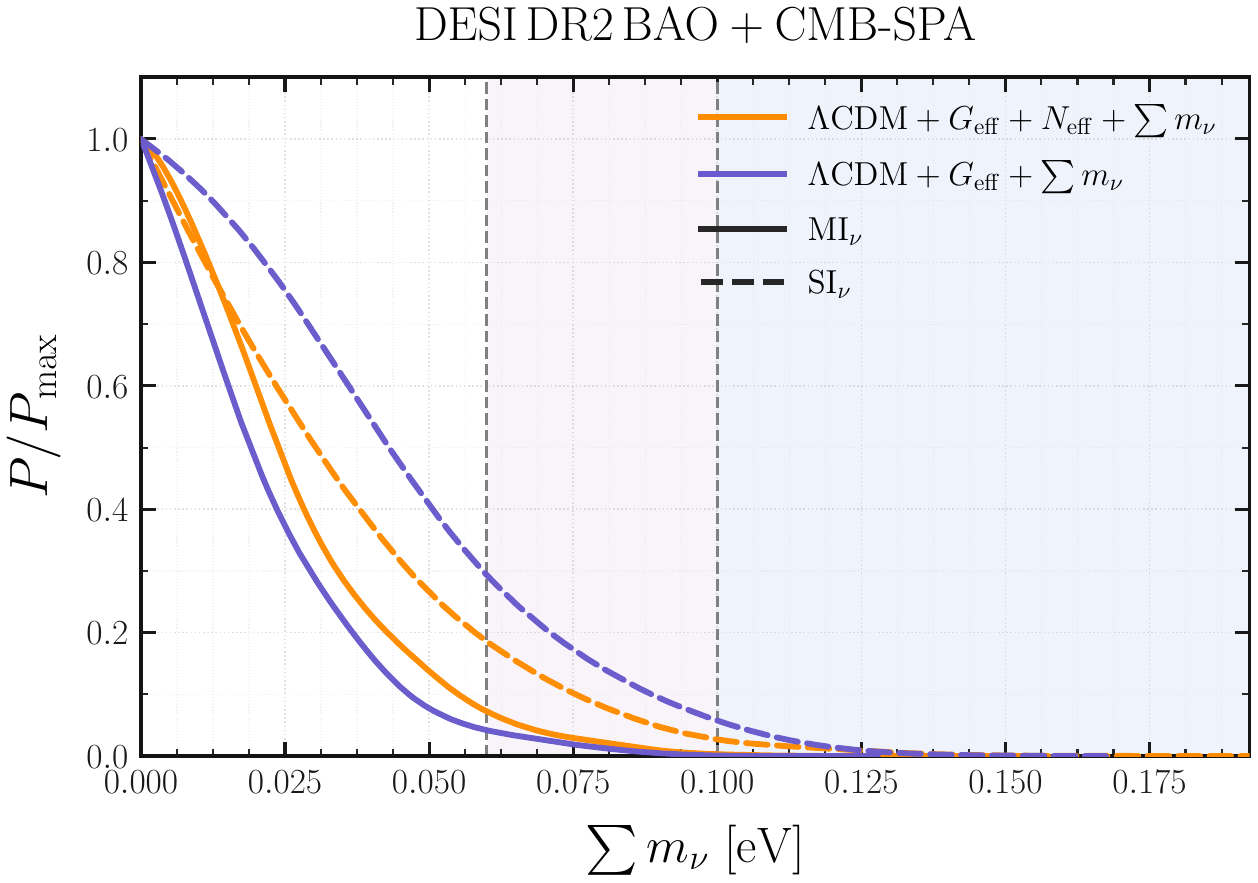}
    }
    \caption{One-dimensional marginalized posterior distributions for $\log_{10}(G_{\rm eff\,} {\rm MeV^{2}})$ (left) and $\sum m_\nu$ (right), obtained by fitting the NSI model to the DESI DR2 BAO+CMB-SPA data. Solid curves represent the moderate mode, while dashed curves correspond to the strong mode.} 
    \label{fig:1d_posterior_cmb_bao}
\end{figure*}

We begin by discussing the results for the DESI DR2 BAO+CMB-SPA data combination.
The reader can see the contours in Fig.~\ref{fig:cmb_bao} and the parameter constraints in Table~\ref{tab:cosmo_cmb}.
The first interesting results with this dataset appear in the MI$_\nu$ mode.
Remarkably, we obtain an anti-correlation between $G_{\rm eff}$ and $n_s$. This is not new and has been spotted since the first NSI WMAP analysis~\cite{CyrRacine2014}, the overall measurement of $n_s$ remains pretty similar to previous CMB results, yet, CMB-SPA has a better resolution to clearly observe this anti-correlation.
This last fact is due to the high-$\ell$ data from SPT and ACT, which provide more sensitivity on $n_s$.
Furthermore, we also spot a positive correlation in the $G_{\rm eff}$--$\sigma_8$ plane. This is expected too, but in our analysis the relation becomes more prominent again.
As the reader may have noticed, in Fig.~\ref{fig:cmb_bao} the bounds on the other cosmological parameters do not exactly match the $\Lambda$CDM results (shown in gray in Fig.~\ref{fig:cmb_bao}).
Notably, this report provides the first CMB-based evidence of small deviation, yet consistent, from standard bounds in the moderate region, which is accompanied by a peak at non-null values in the $G_{\rm eff}$ posterior distribution (this peak is similar to the one observed by ACT DR6~\cite{Calabrese2025atacama}). 

For the SI$_\nu$ regime, the data allow large deviations for most parameters with respect to $\Lambda$CDM.
In particular, $H_0$ shifts toward larger values, yielding $H_0=68.68\pm 0.27$ $\mathrm{km\,s^{-1}\,Mpc^{-1}}$. Although this shift brings the inferred value closer to the {\rm SH$0$ES} determination~\cite{Riess:2021jrx}, a discrepancy of approximately $4.1\sigma$ remains.

Upward shifts in $H_0$ have been reported in CMB analyses, but their magnitude depends on the assumed neutrino-sector parameters and the data combination~\cite{Kreisch2020, Blinov2019, Kreisch:2022zxp}. In particular, the substantially higher values found in some earlier analyses were associated with elevated $N_{\rm eff}$~\cite{Kreisch2020, Blinov2019, Kreisch:2022zxp}.
In Ref.~\cite{Brinckmann2020self, Choudhury2021, Kreisch:2022zxp, Das:2023npl}, it was subsequently found that the SI$_\nu$ regime no longer provided a viable solution to the Hubble tension once high-$\ell$ \textit{Planck 2018} polarization was included.

Continuing with our discussion, we now evaluate the impact of  $\sum m_\nu$ and $N_{\rm eff}$ on the analysis by considering different assumptions for these parameters. 
The  main parameter constraints are summarized in Table~\ref{tab:cosmo_cmb}, while the marginalized posteriors for $\log_{10}(G_{\rm eff}\,{\rm MeV}^{2})$ and $\sum m_\nu$ are shown in Fig.~\ref{fig:1d_posterior_cmb_bao} (for further details see Fig.~\ref{fig:appen_desi_cmb_neff_mnu}).
Focusing first on the NSI parameter in the MI$_\nu$ case, the posterior peak is slightly shifted toward larger values of $\log_{10}(G_{\rm eff},{\rm MeV}^{2})$ and its uncertainty is reduced slightly, we also observe that this is mainly due to free $N_{\rm eff}$, as the neutrino mass is not affecting the bound. Given the fact that the parameter space is extended, initially we did not expect this to be the case, with arguably one explanation being that the data is preferring non-null values.
Furthermore, this is similar to what we observed in the previous section (LSS-only data) when using $\sum m_\nu$ free, although the improvement is not very significant. 

For SI$_\nu$, the peak shifts to lower values. Again, this effect is due mainly to $N_{\rm eff}$.
Thus, this mode appears to peak at $\log_{10}(G_{\rm eff,} {\rm MeV^{2}})= -2$ (the boundary where the priors were separated).
We conclude two things here: one is that, opposite to $N_{\rm eff}$, the neutrino mass is not affecting the $G_{\rm eff}$ bound, and also that for both free standard parameters our prior choice is probably not ideal because the SI$_{\nu}$ mode may suffer from boundary effects.

The corresponding one-dimensional marginalized posteriors for the total neutrino mass, shown in the right panel of Fig.~\ref{fig:1d_posterior_cmb_bao}, illustrate how a delayed onset of neutrino free streaming can relax the bounds on $\sum m_\nu$, alleviating the apparent tension with the minimum total mass implied by neutrino oscillations. For fixed $N_{\rm eff}=3.046$, the 95\% upper limit increases from $\sum m_\nu<0.0509\,\mathrm{eV}$ in the $\mathrm{MI}_\nu$ mode to $<0.0837\,\mathrm{eV}$ in the $\mathrm{SI}_\nu$ mode, around $60\%$ relaxation. With $N_{\rm eff}$ free, the corresponding limits are $<0.0560\,\mathrm{eV}$ and $<0.0768\,\mathrm{eV}$, yielding a $37\%$ relaxation. Compared with the case of fixed $N_{\rm eff}$, allowing this parameter to vary mildly weakens the $\mathrm{MI}_\nu$ mass bound but tightens the $\mathrm{SI}_\nu$ bound, with the respective $N_{\rm eff}$ posteriors centered at $3.15$ and $2.94$. These shifts are associated with the parameter degeneracies involving $N_{\rm eff}$ and $\sum m_\nu$.

\begin{figure*}
    \centering
    \includegraphics[width=1\linewidth]{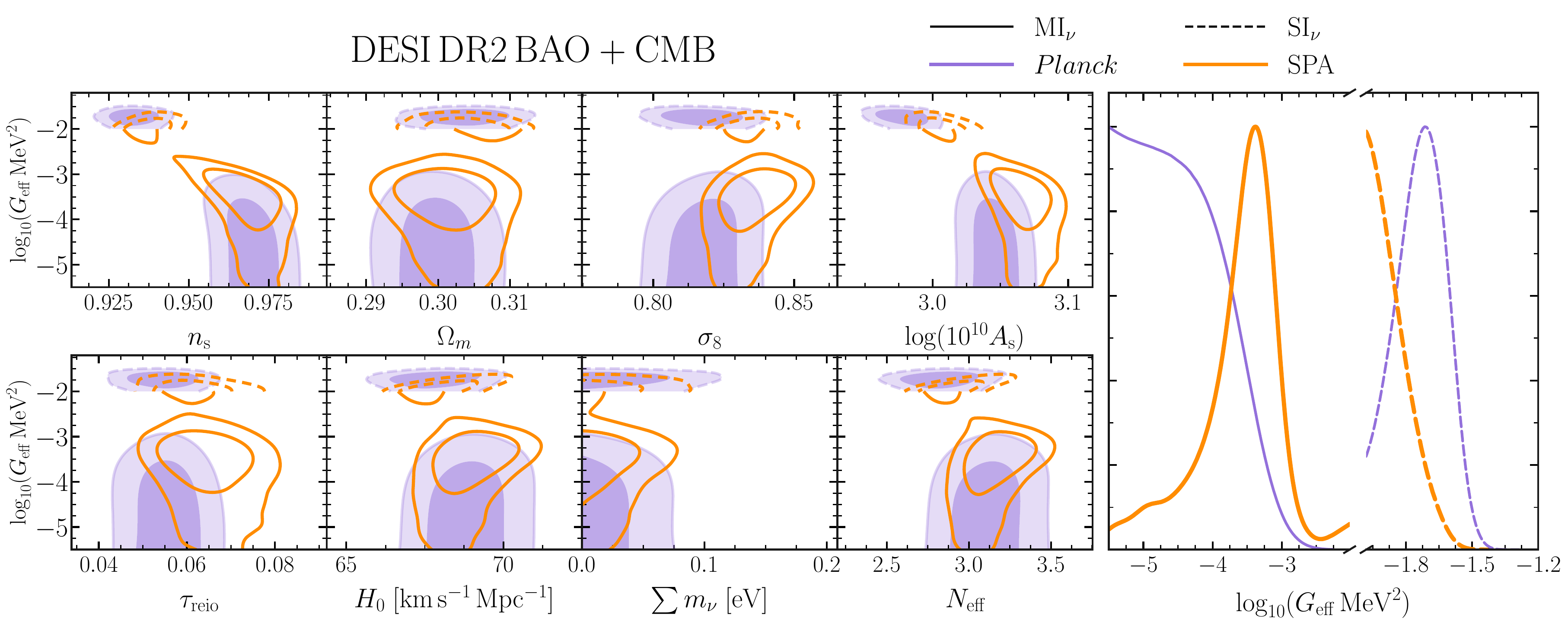}
    \caption{Comparison of the one- and two-dimensional posterior distributions for $\log_{10}(G_{\rm eff,} {\rm MeV^{2}})$ in the NSI model, obtained from two CMB datasets: \textit{Planck 2018} and CMB-SPA.} 
    \label{fig:cmb_bao_planck_vs_spa}
\end{figure*}

Finally, we assess the role of ACT and SPT in the preference for nonzero NSI in the moderately interacting regime. To this end, we perform an additional MCMC analysis using Planck CMB data combined with DESI DR2 BAO. The resulting constraints are presented in Table~\ref{tab:cosmo_cmb} and Fig.~\ref{fig:cmb_bao_planck_vs_spa}. Although the two data combinations yield broadly consistent constraints, the preference for nonzero self-interactions appears only when ACT and SPT are included. This comparison suggests that the additional CMB information provided by ACT and SPT plays a key role in the preference reported above.
Furthermore, including ACT and SPT show more clearly its correlations with $n_s$, $A_s$, and $\sigma_8$.

The strongly interacting regime also changes appreciably. With Planck, the posterior exhibits a distinct maximum near $\log_{10}(G_{\rm eff},\mathrm{MeV}^{2})\simeq-1.7$. With CMB-SPA, it instead rises toward the lower boundary of the strongly interacting prior. Thus, including ACT and SPT favors smaller interaction strengths within this regime.

\subsection{\label{sec:results_cmb_fs} DESI DR1 (FS+BAO) and CMB-SPA joint analysis}

\begin{figure}
    \centering
    \includegraphics[width=1\linewidth]{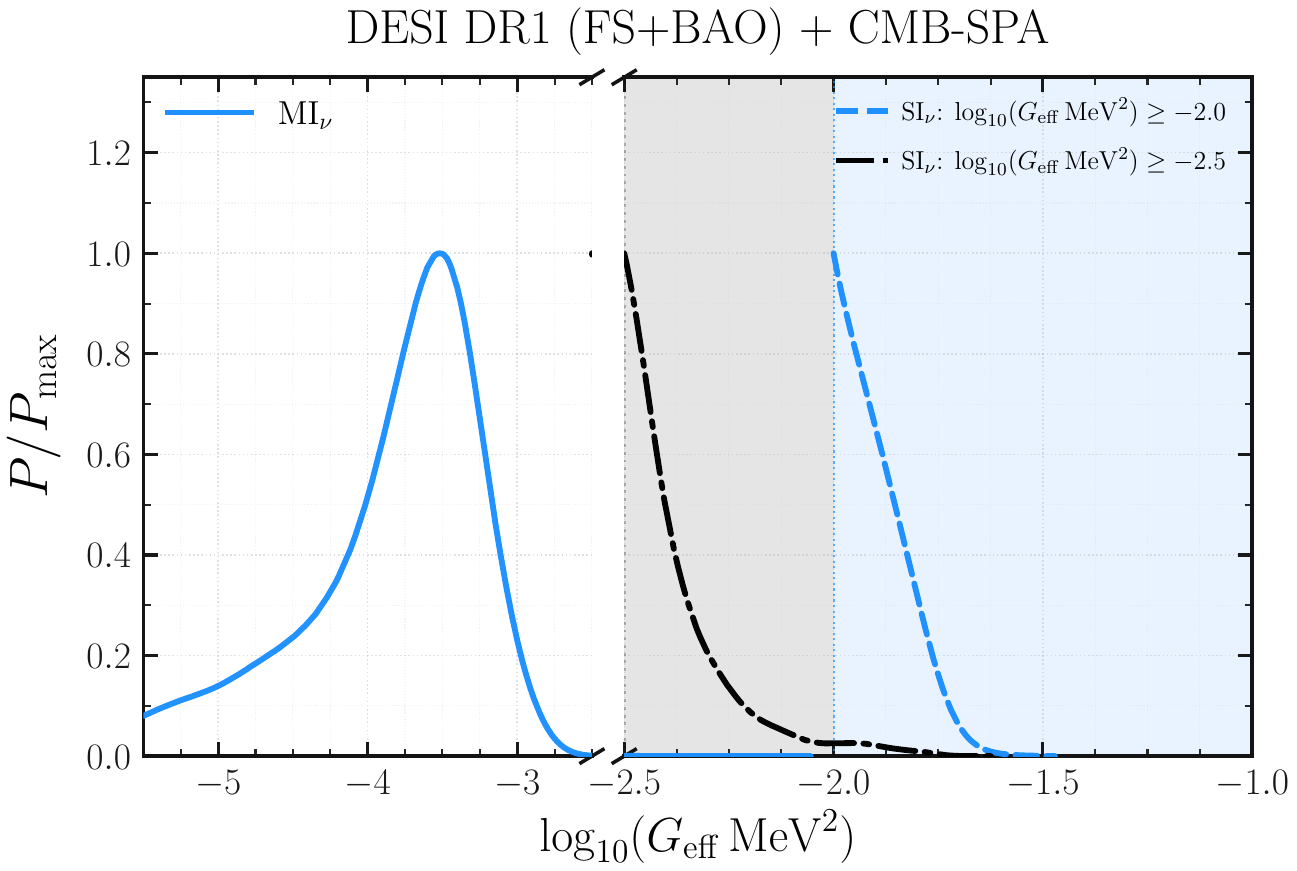}
    \caption{1-D  posterior distributions of $\log_{10}(G_{\rm eff}\,{\rm MeV}^{2})$ obtained from DESI DR1 (FS+BAO)+CMB-SPA, each normalized to its maximum. The blue solid and dashed curves show the moderately interacting and strongly interacting regimes, respectively. The black dash-dotted curve shows the posterior obtained by extending the lower bound of the prior for the strongly interacting regime from $-2.0$ to $-2.5$. The posterior maximum follows the lower prior boundary, indicating that the apparent peak at $\log_{10}(G_{\rm eff}\,{\rm MeV}^{2})=-2.0$ is a boundary effect. Cosmological constraints are reported in Table~\ref{tab:cosmo_cmb}, while the full contour plot is presented in Fig.~\ref{fig:appen_cmb_and_desidr1_cmb}.
    }
    \label{fig:1d_posterior_fs_bao_cmb}
\end{figure}

Finally, we consider the CMB-SPA+DESI DR1 (FS+BAO) data combination, which we expect to provide the most stringent test on NSI among the combinations studied here, owing to the complementary signatures of NSI in the CMB temperature and polarization spectra and the matter power spectrum. The results are summarized in \ref{tab:cosmo_cmb} and in Figs. \ref{fig:1d_posterior_fs_bao_cmb} and \ref{fig:appen_cmb_and_desidr1_cmb}.
Our main result here is that the strong mode disappears from the posterior, with the MCMC favoring values below the prior range.
In order to support our claim, we discard the boundary effect by also running an MCMC with a different prior  $\log_{10}(G_{\rm eff}\, {\rm MeV^{2}})> -2.5$ (see black posterior in Fig.~\ref{fig:1d_posterior_fs_bao_cmb}), and observe that, in fact, the data are not accepting high values of $G_{\rm eff}$.
We can verify that the disappearance of this mode comes from the addition of DESI DR1 (FS+BAO), as shown in Fig.~\ref{fig:appen_cmb_and_desidr1_cmb}.
Nevertheless, we can quantify how much the other parameters changed in this region, overall, we spot cosmological results similar to those reported in the previous subsection, which redundantly represent a divergent cosmology compared to the MI$_\nu$ mode. 
As for the moderate prior, the inclusion of FS maintains consistency of the results (in contrast to the strong mode), giving us similar bounds that in the previous dataset.
We also observe the same correlations between $G_{\rm eff}$ and cosmological standard parameters such as $n_s$ and $A_s$.

As a concluding remark of this section, 
Figs \ref{fig:1d_posterior_cmb_bao} and \ref{fig:1d_posterior_fs_bao_cmb} exhibit maximums of the probability
for the moderate model at $\log_{10}(G_{\rm eff\,} {\rm MeV^{2}})$ close to $-3.5$, this would be interpreted as
a detection of an interaction in an intermediate regime; however, we should be careful. 
First we see that the probability on $G_{\rm eff\,}$ is
$P(G_{\rm eff\,}) \propto  P(\log_{10}(G_{\rm eff\,} {\rm MeV^{2}}))/G_{\rm eff\,}$
and the $1/G_{\rm eff\,}$ 
suppresses the maximum. This result, however, was obtained assuming a flat prior in $\log_{10}(G_{\rm eff\,} {\rm MeV^{2}})$; if we had assumed a flat prior in $G_{\rm eff\,}$, the $1/G_{\rm eff\,}$ cancels and we get 
$P_{\rm{Flat G}}(G_{\rm eff\,}) =  P_{\rm{Flat log}}(\log_{10}(G_{\rm eff\,}))$ which has a local maxima.
For the BAO DR2 + CMB-SPA dataset with fixed ${\sum m_{\nu}}$ and ${N_{\rm eff}}$ the maximum has low
significance and we obtain the bound of $G_{\rm eff } < 0.00308\,\mathrm{MeV}^{-2}$; meanwhile for the same
data but with free $\sum m_\nu$ and $N_{\rm eff}$ we obtain a measurement of $G_{\rm eff} = (8.2^{+10}_{-8.0})\times 10^{-4}\mathrm{MeV}^{-2}$ both reported at 95\% CL (which we argue is not significant enough to report it as evidence of nonstandard NSI).
As for the joint (FS+BAO) DR1+CMB-SPA data we obtain a bound at 95\% CL of $G_{\rm eff} < 0.00145\, \mathrm{MeV}^{-2}$.
In order to probe the viability of each mode, in the next section, we will compute the Bayesian evidence of each mode for the four main datasets considered in this analysis.

\begin{table*}
\caption{Parameter constraints from DESI+CMB datasets.
Parameter uncertainties are reported at 68\% CL and upper bounds are
reported at 95\% CL. The Hubble constant $H_0$ is quoted in units of
$\mathrm{km\,s^{-1}\,Mpc^{-1}}$.}
\label{tab:cosmo_cmb}

\centering

\begingroup
\renewcommand{\arraystretch}{1.20}
\setlength{\tabcolsep}{3pt}

\resizebox{0.98\textwidth}{!}{%
\begin{tabular}{lccccccc}
\hline\hline

{\footnotesize Model/Dataset} &
{\footnotesize $\log_{10}(G_{\rm eff}\,{\rm MeV}^{2})$} &
{\footnotesize $H_0$} &
{\footnotesize $\Omega_m$} &
{\footnotesize $\sigma_8$} &
{\footnotesize $n_s$} &
{\footnotesize $\sum m_\nu$ [$\mathrm{eV}$]} &
{\footnotesize $N_{\rm eff}$}
\\[0.5ex]

\hline


\multicolumn{8}{l}{
  \rule{0pt}{2.8ex}
  $\bm{\Lambda}$\textbf{CDM}
} \\[0.5ex]

\hspace{1em}{\footnotesize DESI DR2 BAO+CMB-SPA}
  & ---
  & $68.24\pm0.25$
  & $0.3027\pm0.0034$
  & $0.8160\pm0.0041$
  & $0.9736\pm0.0030$
  & ---
  & ---
  \\[0.9ex]

\hline


\multicolumn{8}{l}{
  \rule{0pt}{2.8ex}
  $\bm{\Lambda}$\textbf{CDM}+$\bm{G_{\rm eff}}$
} \\[0.5ex]

\hspace{1em}{\footnotesize DESI DR2 BAO+CMB-SPA}
  & & & & & & & \\[0.5ex]

\hspace{1.5em}{\footnotesize MI$_\nu$}
  & $-3.81^{+0.80}_{-0.45}$
  & $68.07\pm 0.26$
  & $0.3050\pm 0.0035$
  & $0.8227^{+0.0044}_{-0.0050}$
  & $0.9663^{+0.0086}_{-0.0041}$
  & ---
  & ---
  \\[0.4ex]

\hspace{1.5em}{\footnotesize SI$_\nu$}
  & $-1.820\pm 0.091$
  & $68.68\pm0.27$
  & $0.3043\pm0.0037$
  & $0.8333\pm0.0047$
  & $0.9419\pm0.0034$
  & ---
  & ---
  \\[0.9ex]

\hspace{1em}{\footnotesize
\shortstack[l]{
  DESI DR1 (FS+BAO)\\
  +CMB-SPA
}}
  & & & & & & & \\[0.5ex]

\hspace{1.5em}{\footnotesize MI$_\nu$}
  & $-3.80^{+0.68}_{-0.29}$
  & $67.64\pm0.32$
  & $0.3110\pm0.0044$
  & $0.8221\pm0.0044$
  & $0.9650^{+0.0067}_{-0.0049}$
  & ---
  & ---
  \\[0.4ex]

\hspace{1.5em}{\footnotesize SI$_\nu$}
  & $< -1.75$
  & $68.65\pm0.32$
  & $0.3043\pm0.0044$
  & $0.8344^{+0.0042}_{-0.0047}$
  & $0.9425\pm0.0033$
  & ---
  & ---
  \\[0.9ex]

\hline

\multicolumn{8}{l}{
  \rule{0pt}{2.8ex}
  $\bm{\Lambda}$\textbf{CDM}+$\bm{G_{\rm eff}}$
} \\[-0.1ex]

\multicolumn{8}{l}{
  \hspace{1.5em}
  $+\bm{\sum m_\nu}$
} \\[0.5ex]

\hspace{1em}{\footnotesize DESI DR2 BAO+CMB-SPA}
  & & & & & & & \\[0.5ex]

\hspace{1.5em}{\footnotesize MI$_\nu$}
  & $-3.93^{+0.81}_{-0.34}$
  & $68.32\pm 0.26$
  & $0.3027\pm 0.0035$
  & $0.8302\pm 0.0051$
  & $0.9669^{+0.0070}_{-0.0043}$
  & $< 0.0509$
  & ---
  \\[0.4ex]

\hspace{1.5em}{\footnotesize SI$_\nu$}
  & $-1.843^{+0.078}_{-0.11}$
  & $68.87^{+0.31}_{-0.28}$
  & $0.3024\pm 0.0038$
  & $0.8389^{+0.0071}_{-0.0053}$
  & $0.9416\pm 0.0034$
  & $< 0.0837$
  & ---
  \\[0.9ex]

  \hline

\multicolumn{8}{l}{
  \rule{0pt}{2.8ex}
  $\bm{\Lambda}$\textbf{CDM}+$\bm{G_{\rm eff}}$
} \\[-0.1ex]

\multicolumn{8}{l}{
  \hspace{1.5em}
  $+\bm{\sum m_\nu}+\bm{N_{\rm eff}}$
} \\[0.5ex]

\hspace{1em}{\footnotesize DESI DR2 BAO+CMB-SPA}
  & & & & & & & \\[0.5ex]

\hspace{1.5em}{\footnotesize MI$_\nu$}
  & $-3.61^{+0.61}_{-0.30}$
  & $68.91\pm 0.82$
  & $0.3015\pm 0.0041$
  & $0.8353\pm 0.0078$
  & $0.9675^{+0.0080}_{-0.0054}$
  & $< 0.0560$
  & $3.15\pm 0.13$
  \\[0.4ex]

\hspace{1.5em}{\footnotesize SI$_\nu$}
  & $< -1.69$
  & $68.24^{+0.73}_{-0.93}$
  & $0.3037\pm0.0039$
  & $0.8351\pm0.0073$
  & $0.9383\pm0.0048$
  & $< 0.0768$
  & $2.94^{+0.11}_{-0.15}$
  \\[0.9ex]

\hspace{1em}{\footnotesize
\shortstack[l]{
  DESI DR2 BAO\\
  +\textit{Planck}
}}
  & & & & & & & \\[0.5ex]

\hspace{1.5em}{\footnotesize MI$_\nu$}
  & $<-3.38$
  & $68.88\pm0.94$
  & $0.2998\pm0.0041$
  & $0.8184\pm0.0096$
  & $0.9693\pm0.0063$
  &$< 0.0669$
  & $3.12^{+0.15}_{-0.17}$
  \\[0.4ex]

\hspace{1.5em}{\footnotesize SI$_\nu$}
  & $-1.74^{+0.13}_{-0.10}$
  & $67.85^{+0.90}_{-1.0}$
  & $0.3037\pm0.0042$
  & $0.8174\pm0.0098$
  & $0.9329\pm0.0053$
  & $< 0.0981$
  & $2.83^{+0.15}_{-0.17}$
  \\[0.9ex]

\hline\hline
\end{tabular}%
}

\endgroup
\end{table*}

\section{\label{sec:bayesian} Mode Comparison}

In this section, we compare the relative statistical support for the
two modes within the NSI framework using the Bayes factor and the
maximum-likelihood ratio. 
For a model $\mathcal{M}$ with parameters $\boldsymbol{\theta}$, the Bayesian evidence is given by (see, e.g.,~\cite{Jaynes2003probability, Mackay2003information})
    \begin{equation}
        \mathcal{Z}= \int \mathcal{L}(\mathbf{D}|\boldsymbol{\theta}, \mathcal{M})\, \pi(\boldsymbol{\theta}|\mathcal{M})\, d\boldsymbol{\theta},
    \end{equation}
where $\mathbf{D}$ is the dataset, $\mathcal{L}$ is the likelihood and $\pi(\boldsymbol{\theta}|\mathcal{M})$ is the prior distribution. 
To compare the modes, we evaluate their evidences separately, with priors normalized within their respective parameter regions. The ratio of the evidences of modes $i$ and $j$ defines the
Bayes factor,
\begin{equation}
B_{ij} \equiv \frac{\mathcal{Z}_i}{\mathcal{Z}_j}, \qquad \ln B_{ij} = \ln \mathcal{Z}_i - \ln \mathcal{Z}_j.
\end{equation}
Here, we use the Bayes factor to compare two modes of the posterior within the same NSI model.

For this purpose, we use the code \textsc{harmonic}\footnote{\gitlinkharmonic} which computes the Bayes factor relying on machine learning algorithms to train an optimal target distribution directly from posterior samples~\cite{harminocmcewen2021machine}. 
To assess the robustness of our results, we performed a few validation runs with \textsc{pocomc}\footnote{\gitlinkpocomc}~\cite{pocomc1Karamis, Karamanis2022pocomc}. We found good agreement between the two codes.

Since we compare the SI$_\nu$ and MI$_\nu$ regimes, we use $\ln B_{\rm S_\nu M_\nu}$, with the positive (negative) values favoring the strong (moderate) mode.
We show our results in Table~\ref{table:bayes}. For reference, we interpret the results using the Jeffreys scale~\cite{Jeffreys1939-JEFTOP-5, Kass1995}: $\left|\ln B_{\rm S_\nu M_\nu}\right|<1$ is considered inconclusive, $1\leq\left|\ln B_{\rm S_\nu M_\nu}\right|<2.5$ as weak evidence, $2.5\leq\left|\ln B_{\rm S_\nu M_\nu}\right|<5$ as moderate evidence, and  $\left|\ln B_{\rm S_\nu M_\nu}\right|>5$ as strong evidence. 

We also compare the two modes through their maximum-likelihood ratio,
\begin{equation}
    R_{\rm S_\nu M_\nu}
    =
    \frac{\max(\mathcal{L}_{\rm SI_\nu})}
         {\max(\mathcal{L}_{\rm MI_\nu})},
\end{equation}
where $\mathcal{L}_{\rm SI_\nu}$ and
$\mathcal{L}_{\rm MI_\nu}$ denote the likelihood restricted
to the parameter regions associated with the SI$_\nu$ and
MI$_\nu$ modes, respectively.
This ratio complements the Bayes factor by comparing the
best-fit likelihoods, whereas the evidence averages the
likelihood over the adopted prior.

In Table~\ref{table:bayes}, we observe that, when using LSS-only data, the evidence slightly favors the MI$_\nu$ mode, and becomes moderate with the inclusion of Ly-$\alpha$ data.
However, we recall that the MI$_\nu$ region is bimodal with Ly-$\alpha$, thus, we proceed to compare the WI$_\nu$ and the II$_\nu$ modes within this region (separated at $\log_{10}(G_{\rm eff\,} {\rm MeV^{2}})=-3.735$). In this case, we found the Bayes factor to be $\ln B_{\rm W_\nu I_\nu}=-0.28\pm0.36$, which we interpret as neither mode is favored. 

For the BAO DR2+CMB-SPA joint dataset, the evidence against the SI$_\nu$ mode is moderate.
Here, we note two points: first, including ACT and SPT strengthens
the evidence compared to \textit{Planck}-only CMB data under the
same neutrino-parameter assumptions, consistent with the MCMC
results; second, the evidence is stronger when the standard
neutrino parameters are allowed to vary than when they are fixed.

Furthermore, for the DESI DR1 (FS+BAO)+CMB-SPA dataset, we obtain
$\ln B_{\rm S_\nu M_\nu}=-9.91\pm0.10$, indicating very strong
evidence against the SI$_\nu$ mode. This corresponds to a posterior probability of approximately $5$ in $100000$ for the SI$_\nu$ mode.
This is not surprising, since we saw in the MCMC analysis that the strong mode is heavily suppressed. Our result reinforces previous evidence against this mode~\cite{Camarena2025strong, Poudou:2025qcx}.
Together, these findings further constrain the viability of
the SI$_\nu$ regime.

As a complementary measure, we report the maximum-likelihood ratios in Table~\ref{table:bayes}. For DESI DR1 (FS+BAO)+BBN+$n_{s10}$, we obtain $R_{\rm S_\nu M_\nu}=1.44$, indicating that the SI$_\nu$ region has a slightly higher maximum likelihood than the MI$_\nu$ region. Nevertheless, the Bayesian evidence favors MI$_\nu$, with $\ln B_{\rm S_\nu M_\nu}=-2.77\pm0.38$. 

For BAO DR2+CMB-SPA with both $N_{\rm eff}$ and $\sum m_\nu$ free, the Bayes factor and the maximum-likelihood ratio both favor the MI$_\nu$ mode. For DESI DR1 (FS+BAO)+CMB-SPA, the maximum-likelihood ratio is particularly small, $R_{\rm S_\nu M_\nu}=4.53\times10^{-4}$, indicating that the SI$_\nu$ region provides a substantially poorer best fit than the MI$_\nu$ region.

\begin{table*}[htb]
\caption{
Bayes factors and maximum likelihood ratios for different datasets.
Unless indicated otherwise, $N_{\rm eff}=3.046$ and
$\sum m_\nu=0.06\,\mathrm{eV}$ are fixed. For $\ln B_{\rm S_\nu M_\nu}$ positive (negative) values favor the strongly (moderately)
interacting mode.
Values of the maximum likelihood ratio $R_{\rm S_\nu M_\nu}$ above (below) unity favor the strongly (moderately) interacting mode.
The $\ln B_{\rm W_\nu I_\nu}$ and $R_{\rm W_\nu I_\nu}$ comparisons use
the modes separated at
$\log_{10}(G_{\rm eff}\,\mathrm{MeV}^{2})=-3.735$
with Ly-$\alpha$ data.
}
\label{table:bayes}

\begingroup
\renewcommand{\arraystretch}{1.15}
\setlength{\tabcolsep}{4pt}

\begin{ruledtabular}
\begin{tabular}{lrr}
Dataset / assumptions & $\ln B_{\rm S_\nu M_\nu}$ & $R_{\rm S_\nu M_\nu}$ \\
\hline
\noalign{\vskip 3pt}
\shortstack[l]{DESI DR1 (FS+BAO)\\
\quad + BBN + $n_{s10}$}
& $-2.77 \pm 0.38$ & $1.44$ \\
\noalign{\vskip 3pt}
\shortstack[l]{DESI DR1 (FS+BAO+Ly-$\alpha$)\\
\quad + BBN + $n_{s10}$}
& $-3.52 \pm 0.33$ & $0.0875$ \\
\multicolumn{3}{c}{\dotfill} \\
CMB-SPA + DESI DR1 (FS+BAO)
& $-9.91 \pm 0.10$ & $4.53 \times 10^{-4}$ \\
\multicolumn{3}{c}{\dotfill} \\
DESI DR2 BAO + CMB-SPA & & \\
\quad Both fixed
& $-1.95 \pm 0.02$ & $0.472$ \\
\quad $\sum m_\nu$ free
& $-3.54 \pm 0.03$ & $0.0636$ \\
\quad $N_{\rm eff}$ and $\sum m_\nu$ free
& $-3.37 \pm 0.02$ & $0.123$ \\
\multicolumn{3}{c}{\dotfill} \\
DESI DR2 BAO + \textit{Planck} & & \\
\quad $N_{\rm eff}$ and $\sum m_\nu$ free
& $-1.26 \pm 0.04$ & $0.536$ \\
\hline\hline
Dataset & $\ln B_{\rm W_\nu I_\nu}$ & $R_{\rm W_\nu I_\nu}$ \\
\hline
\noalign{\vskip 3pt}
\shortstack[l]{DESI DR1 (FS+BAO+Ly-$\alpha$)\\
\quad + BBN + $n_{s10}$}
& $-0.28 \pm 0.36$ & $0.687$ \\
\end{tabular}
\end{ruledtabular}

\endgroup
\end{table*}

\section{\label{sec:conclusions}Summary and conclusions}

In this work, we investigated the simplest scenario of neutrino self-interactions in the heavy-mediator limit, parameterized by a single additional coupling, $G_{\rm eff}$. We analyzed DESI DR1 full-shape galaxy clustering and BAO measurements, with and without Ly-$\alpha$ constraints, supplemented by BBN and a weak prior on $n_s$. We also combined DESI DR2 BAO or DESI DR1 FS+BAO with CMB-SPA, comprising Planck, ACT, and SPT observations, to exploit the complementarity between large-scale structure and CMB measurements. We characterized the interaction regimes through profile likelihoods, posterior sampling, Bayesian evidence, and maximum-likelihood comparisons.

The Bayesian evidence favors the MI$_\nu$ region over SI$_\nu$ in every comparison performed, although the strength of this preference depends on the dataset and neutrino-sector assumptions. The strongest preference arises from DESI DR1 (FS+BAO)+CMB-SPA, for which we obtain $\ln B_{\mathrm{S_\nu M_\nu}}=-9.91\pm0.10$, corresponding to odds of approximately $5:100000$ for SI$_\nu$ relative to MI$_\nu$. This provides strong Bayesian evidence against the strongly interacting neutrino scenario, in which the onset of neutrino free streaming is delayed until around matter–radiation equality. The maximum-likelihood ratio, $R_{\mathrm{S_\nu M_\nu}}=4.53\times10^{-4}$, independently shows that the SI$_\nu$ region provides a substantially poorer best fit. Moreover, extending its prior toward smaller couplings causes the posterior maximum to follow the lower boundary rather than revealing a distinct strongly interacting solution.

For the combinations without a CMB likelihood, the evidence moderately favors MI$_\nu$. However, DESI DR1 (FS+BAO)+BBN+$n_{s10}$ retains the strongly interacting solution with a slightly higher maximum likelihood than the moderate branch. For DESI DR2 BAO+CMB-SPA, the evidence against SI$_\nu$ ranges from weak to moderate across the neutrino-sector assumptions explored. Under matched assumptions, including ACT and SPT strengthens this preference relative to Planck alone.

The LSS-only analysis also reveals multimodal features. Without Ly-$\alpha$, the profile likelihood admits the moderate and a strongly interacting solution, neither providing an appreciable improvement over $\Lambda$CDM. Including Ly-$\alpha$ resolves an additional narrow mode within the moderate region, which we designate II$_\nu$. For this branch, we obtain $\log_{10}(G_{\rm eff}\,\mathrm{MeV}^{2})=-3.04^{+0.11}_{-0.078}$ at 68\% CL. Its evidence is comparable to that of the weaker, standard-like WI$_\nu$ branch, with $\ln B_{\rm W_\nu I_\nu} =-0.28\pm0.36$. This comparison is inconclusive, and the intermediate solution does not provide an appreciable improvement in the best fit over $\Lambda$CDM.

Future DESI releases, DESI-II, and complementary photometric surveys will improve the sensitivity to $G_{\rm eff}$~\cite{Desi2whitepaper, LSST_Ivezic_2019, DESC_mandelbaum2018, Euclid2024}. Dedicated analyses of the neutrino-induced acoustic phase shift in CMB and BAO observations offer another route to testing delayed neutrino free-streaming~\cite{Montefalcone:2025ibh, Whitford:2025dmq}.
New observables, such as the post-reionization 21-cm~\cite{Pal2026neutrino, meerklasscollaboration2026} and the yet-to-be-detected cosmic dawn 21-cm signals~\cite{Libanore2025}, will also help tighten constraints.
On the theoretical side, the model could be refined with more realistic neutrino self-interaction parameterizations, either through mediator-mass-dependent~\cite{Perez2026towards} or flavor-dependent couplings~\cite{Das2021, Mazumdar2022flavour, Das:2023npl, RoyChoudhury2022, Bostan_2024, Barenboim:2026jef}, especially since existing experimental and astrophysical neutrino bounds are in tension with the baseline model~\cite{Gerda2015results, Huang2018, Blum2018neutrinoless, Blinov2019, Krnjaic2020probing, Brdar2020, Deppisch2020, Lyu2021self, Fiorillo2022, Dutta:2023fdt, Wu2024shedding, Demirci2024, Dev2025new, Zhang2025neutrino, Foroughi2026radiative, Foroughi2026precision}.
To evade these bounds or probe dark-sector physics, several works have explored extended models~\cite{De2020dodelson, He2020, Berbig2020, Kelly2020origin, Chichiri2022cosmological, An2023can, Ghosh2024hubble, Foroughi2025enabling, Das2025impostor, Parashari:2026dxo} as well as alternative regimes, such as resonant interactions~\cite{Venzor2023resonant, Noriega2025resonant} or ultralight mediators~\cite{Forastieri:2015paa, Forastieri:2019cuf, Venzor2022, Sharma2026recoupled, Gil-Ocaranza:2026lpq}.

\appendix 

\section{Extended information on cosmological bounds} \label{appendixA}

\begin{figure*}
    \centering
    \includegraphics[width=\textwidth]{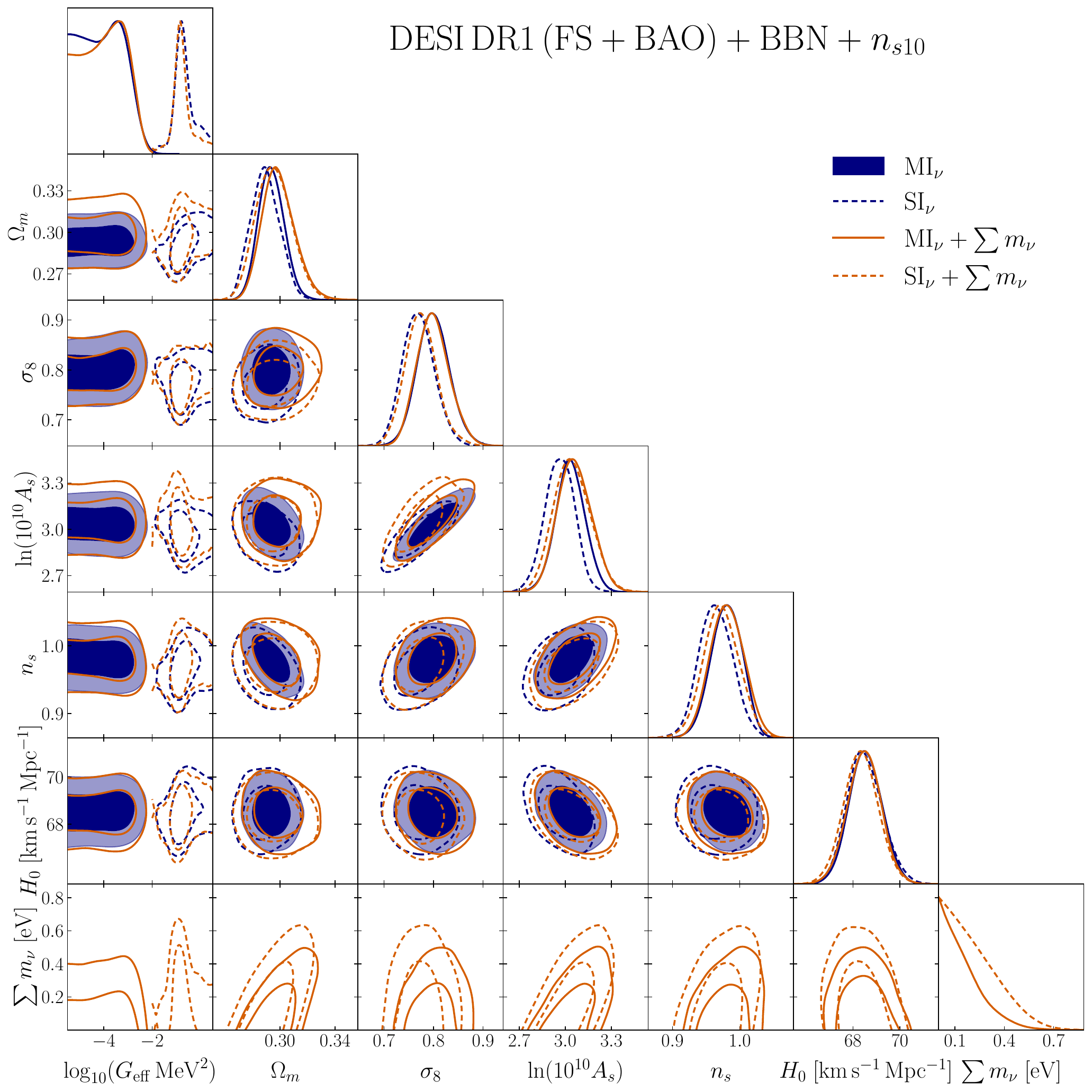}
    \caption{Parameter constraints at 68\% and 95\% CLs for the NSI model with fixed (blue) and free (orange) total neutrino masses, $\sumnu$, using the DESI DR1 (FS+BAO)+BBN+$n_{s10}$ dataset. Solid and dashed curves represent the moderate and strong neutrino self-interaction modes, respectively. 
    }
    \label{fig:appen_fs_mass_nu}
\end{figure*}

\begin{figure*}
    \centering
    \includegraphics[width=\textwidth]{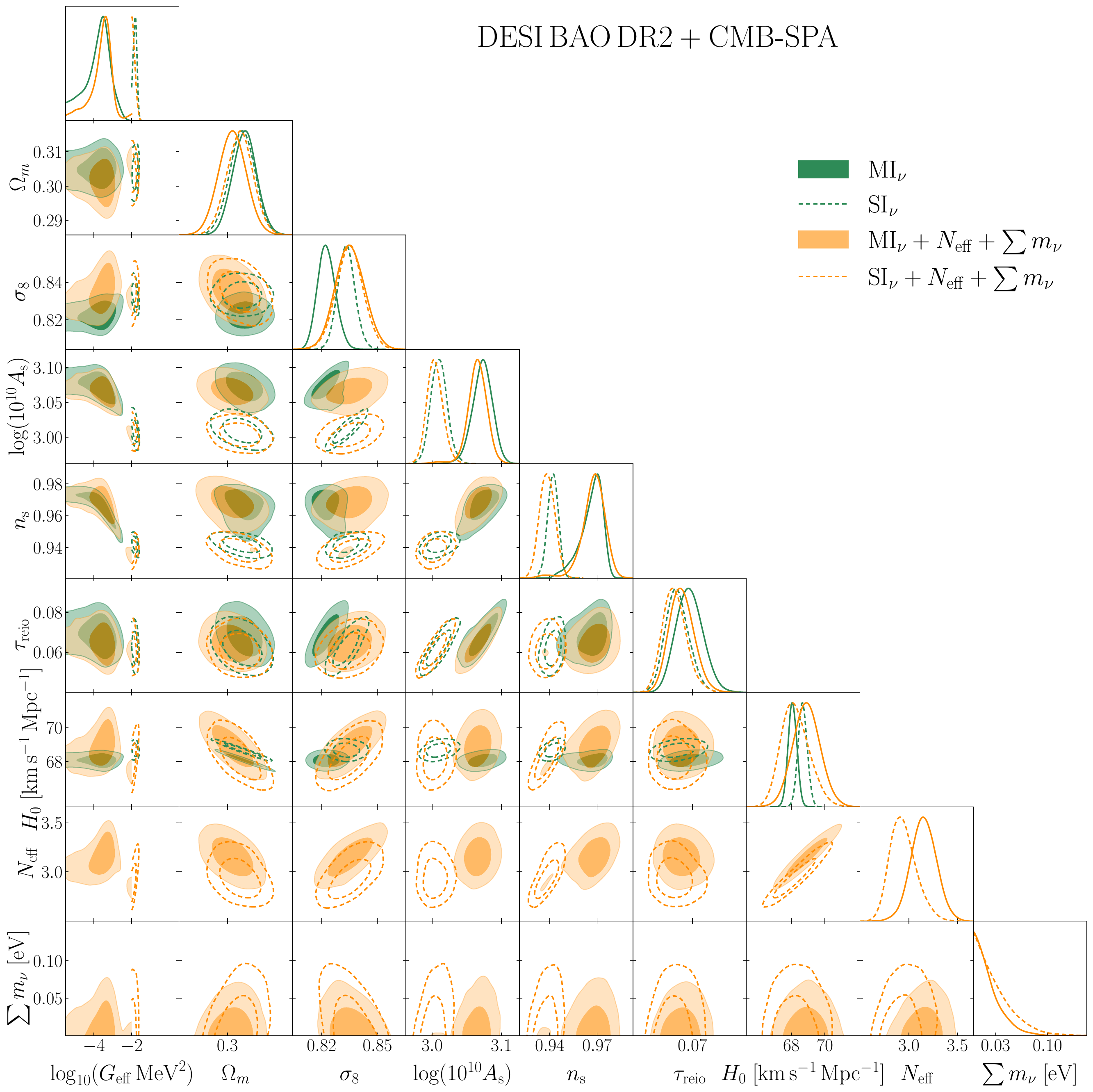}
    \caption{Parameter constraints at 68\% and 95\% CLs for the NSI model with fixed (green) and free (orange) $N_{\rm eff}$ and $\sumnu$, using the DESI DR2 BAO+CMB-SPA dataset. Solid and dashed curves represent the moderate and strong neutrino self-interaction modes, respectively. 
    }
    \label{fig:appen_desi_cmb_neff_mnu}
\end{figure*}

\begin{figure*}
    \centering
    \includegraphics[width=\textwidth]{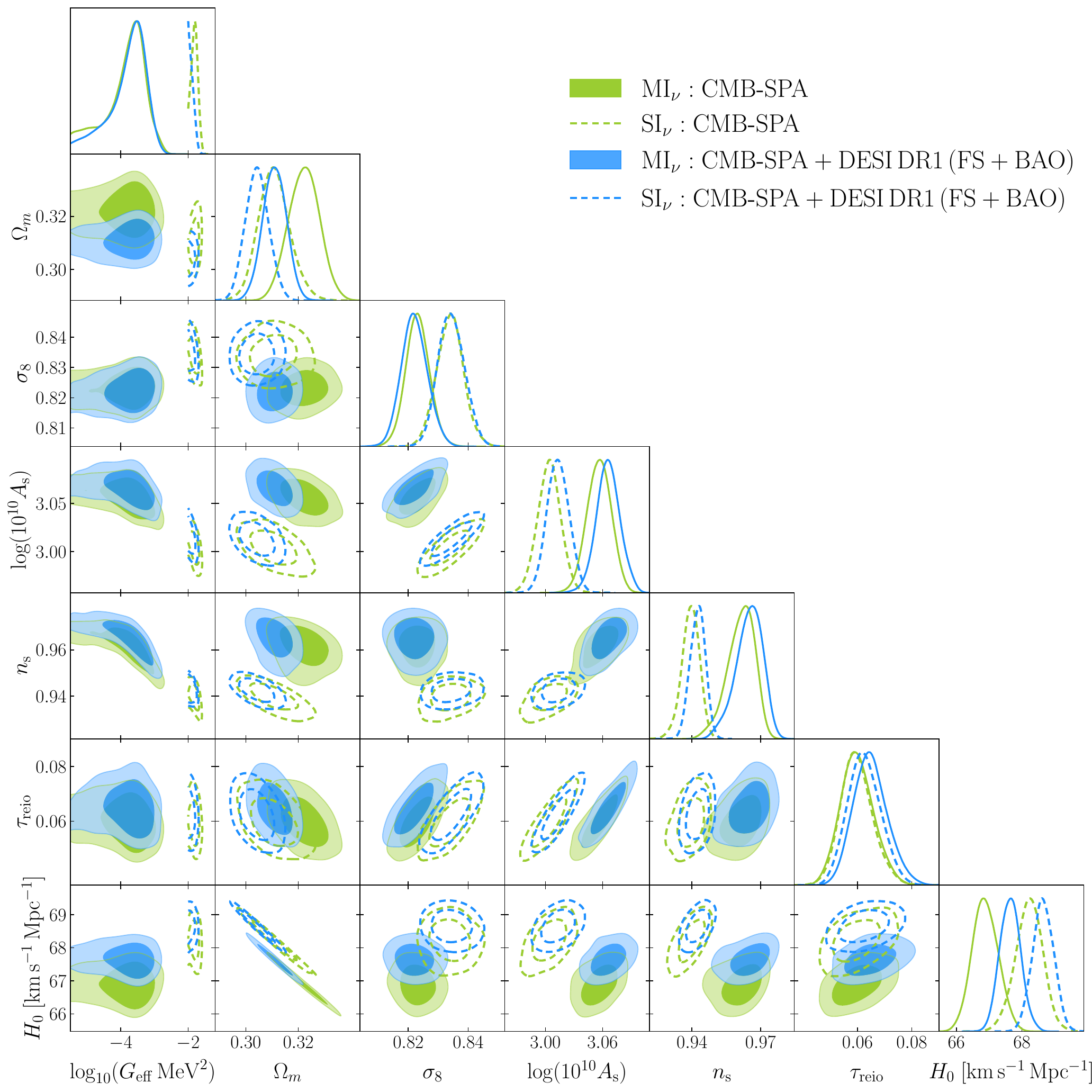}
    \caption{Parameter constraints at 68\% and 95\% CLs for the NSI model, showing the impact of adding DESI DR1 (FS+BAO) data to CMB-SPA. Solid curves and filled contours denote the moderately interacting regime, while dashed curves and unfilled contours denote the strongly interacting regime.
    }
    \label{fig:appen_cmb_and_desidr1_cmb}
\end{figure*}

This appendix presents further contour plots including more cosmological parameters for the different scenarios and datasets considered.

\acknowledgments

H.N. acknowledges support from the Leinweber Institute for Theoretical Physics (LITP) at the University of Michigan Postdoctoral Research Fellowship and support from the Department of Energy under contract DE-SC009193.
G.G.-A. and J.V. acknowledge support from SECIHTI postdoctoral fellowships. 
H.N. acknowledges support from SECIHTI through grants CBF-2025-G-1720 and CBF-2025-G-176.
I.P.C. thanks SECIHTI for the scholarship as ``Ayudante de Investigador SNII III''.
The authors thankfully acknowledge Cl\'uster de Superc\'omputo Xiuhcoatl (Cinvestav) for the allocation of computer resources.

\bibliographystyle{JHEP} 

\bibliography{references}

\end{document}